\documentclass[aps]{revtex4}
\usepackage{eurosym}
\usepackage{amsfonts}
\usepackage{amsmath}
\usepackage{amssymb,epsf}
\usepackage{color}
\usepackage{xcolor}
\usepackage{graphicx}
\usepackage{epstopdf}
\usepackage{float}
\usepackage{caption}
\usepackage{subfig}
\usepackage{hyperref}
\usepackage{orcidlink}

\begin{document}

\title{Thermodynamics, Phase Transitions, and Geodesic Structure of $F(R)-$%
Phantom Banados–Teitelboim–Zanelli (BTZ) Black Holes}
\author{Behzad Eslam Panah\,\orcidlink{0000-0002-1447-3760}}
\email{eslampanah@umz.ac.ir}
\affiliation{Department of Theoretical Physics, Faculty of Basic Sciences, University of
Mazandaran, P. O. Box 47416-95447, Babolsar, Iran}
\author{Bilel Hamil\,\orcidlink{0000-0002-7043-6104}}
\email{hamilbilel@gmail.com/bilel.hamil@umc.edu.dz}
\affiliation{Laboratoire de Physique Math\'{e}matique et Physique Subatomique,LPMPS,
Facult\'{e} des Sciences Exactes, Universit\'{e} Constantine 1, Constantine,
Algeria}
\author{Manuel E. Rodrigues\,\orcidlink{0000-0001-8586-0285}}
\email{esialg@gmail.com}
\affiliation{Faculdade de F\'{\i}sica, Programa de P\'{o}s-Gradua\c{c}\~{a}o em F\'{\i}
sica, Universidade Federal do Par\'{a}, 66075-110, Bel\'{e}m, Par\'{a},
Brazill}
\affiliation{Faculdade de Ci\^{e}ncias Exatas e Tecnologia, Universidade Federal do Par 
\'{a}, Campus Universit\'{a}rio de Abaetetuba, 68440-000, Abaetetuba, Par 
\'{a}, Brazil}

\begin{abstract}
This paper investigates \textit{phantom} BTZ black holes within the
high-curvature gravity theory framework, specifically using a special case
of \textit{power-Maxwell} theory, which functions as a nonlinear
electrodynamics source called \textit{$F(R)-$conformally invariant Maxwell}
gravity. We examine how the \textit{phantom} or anti-Maxwell field affects
the structure of these black holes and how the theory's parameters influence
their horizon structure. Additionally, we derive the conserved and
thermodynamic potentials associated with these black holes, thereby
establishing their conformance to the foundational first law of
thermodynamics. Next, the stability characteristics---both local and
global---of BTZ black holes endowed with \textit{phantom} and Maxwell fields
are explored under canonical and grand canonical ensemble conditions by
inspecting their heat capacity and Gibbs free energy profiles. This
assessment reveals how the \textit{phantom} field and scalar curvature
affect these stability regions. We then perform a rigorous analytical
verification of the Ehrenfest equations to determine whether the critical
behavior of the \textit{phantom} BTZ black hole corresponds to a
second-order phase transition. Our results demonstrate adherence to both
Ehrenfest relations, thereby confirming the occurrence of a second-order
phase transition within the black hole system concurrent with the critical
point. Furthermore, we explore the geodesic structure of the obtained
solutions to analyze the motion of massive and massless test particles in
the $F(R)$-\textit{phantom} BTZ spacetime. The analysis demonstrates that
stable timelike circular orbits exist only in the \textit{phantom} regime
for negative curvature backgrounds, while the \textit{phantom} configuration
also allows for stable circular photon orbits. These results underscore the
significant influence of the \textit{phantom} field and the $F(R)$
correction on the spacetime geometry and orbital dynamics.
\end{abstract}

\maketitle

\section{Introduction}

Among various nonlinear electrodynamics models, one of the most prominent
frameworks is the \textit{power-Maxwell} theory, in which the Lagrangian
density is introduced as an arbitrary power of the classical Maxwell
Lagrangian. This theory, grounded on previous studies \cite%
{PMI,PMII,PMIII,PMIV,PMV}, remains invariant under conformal transformations
of the metric and the electromagnetic potential, i.e., $g_{\mu\nu}
\rightarrow \Omega^2 g_{\mu\nu}$ and $A_{\mu} \rightarrow A_{\mu}$, where $%
g_{\mu\nu}$ is the spacetime metric and $A_{\mu}$ denotes the gauge field.
In the special case where the power equals one, the model naturally reduces
to the linear Maxwell theory. Moreover, by a proper choice of the power
parameter, the theory becomes \textit{conformally invariant}, leading to the
conformal Maxwell formulation. A notable feature of this framework is its
ability to regularize the divergent behavior of the electric field near
point charges, rendering it physically meaningful after parameter
adjustments \cite{PM1,PM2}.

The $F(R)$ gravity theory \cite%
{F(R)1,F(R)2,F(R)3,F(R)4,F(R)5,F(R)6,F(R)7,F(R)8,F(R)9,F(R)10,F(R)11,F(R)12,F(R)14,F(R)15}
stands as one of the well-known extensions of General Relativity (GR)
proposed to explain the observed cosmic acceleration, which cannot be
accounted for within classical GR. Rich in both cosmological and
astrophysical contexts \cite%
{Mod1,Mod2,Mod3,Mod4,Mod5,Mod7,Mod8,Mod9,Mod10,Mod11,Mod12}, it successfully
reproduces the complete evolution history of the Universe---from the early
inflationary epoch through the radiation-dominated period, up to the current
dark energy era (see Refs. \cite%
{CosFR3,CosFR4,CosFR6,CosFR7,CosFR8,CosFR9,CosFR10,CosFR11,CosFR12,CosFR13},
for more details). Furthermore, the theory is consistent with Newtonian and
post-Newtonian limits and can model cosmic structure formation without
invoking dark matter \cite{CapozzielloI,CapozzielloII}.

Black holes represent a cornerstone in theoretical and observational studies
of gravitation, serving as natural laboratories to probe the deep structure
of spacetime in any gravitational theory. While many solutions of Einstein's
GR can also appear within the $F(R)$ framework, the latter admits families
of non-Einsteinian solutions with distinct physical characteristics.
Identifying these classes of black holes and studying their properties
within $F(R)$ gravity is crucial, albeit challenging, due to the nonlinear
fourth-order nature of the field equations. Their analytical treatment,
especially in the presence of matter, is highly nontrivial. Nevertheless,
numerous exact and approximate black hole solutions have been obtained \cite%
{BHFR1,BHFR2,BHFR3,BHFR4,BHFR5,BHFR6,BHFR7,BHFR8,BHFR9,BHFR10,BHFR11,BHFR12,BHFR13,BHFR14,BHFR15,BHFR17,BHFR18,BHFR19,BHFR20,BHFR21,BHFR22}%
, including those where $F(R)$ gravity couples to nonlinear electrodynamics 
\cite{NoBHFR1,NoBHFR2,NoBHFR3,NoBHFR4,NoBHFR5,NoBHFR6,NoBHFR7,NoBHFR8}.

One of the earliest fundamental black hole solutions in three
dimensions is the BTZ black hole, proposed by Banados, Teitelboim, and Zanelli \cite{BTZ}. This geometry has unique characteristics that
simplify the analysis of gravitational interactions and thermodynamic
behavior. Lower-dimensional models allow for more transparent mathematical
treatments and can provide insights applicable to higher dimensions. The BTZ
black hole, in particular, serves as an excellent platform for studying the
effects of modifications in gravity, such as $F(R)$ theories, due to its
rich structure despite its simplicity. This geometry offers a
straightforward and computationally accessible framework for investigating
gravity in lower dimensions, leading to deeper insights into the quantum
nature of spacetime \cite{Witten2007}. Additionally, the BTZ metric connects
gravitational theories with string theory \cite{Witten1998}. While
four-dimensional gravity is non-renormalizable, its three-dimensional
counterpart is exactly solvable and perturbatively renormalizable, providing
an effective environment for exploring quantum black holes and their
underlying phenomena \cite{3d,3d2,3d3,3d4,3d5,3d6,3d7}. By leveraging the
advantages of three-dimensional models, we can gain clearer insights into
essential concepts such as black hole thermodynamics and phase transitions,
without the complications inherent in higher dimensions.

The concept of \textit{phantom} fields dates back to Einstein and Rosen
(1935) \cite{ER}, who proposed a bridge-like structure based on a
Reissner--Nordstr\"{o}m solution with an imaginary charge $q^2 \rightarrow
-q^2$, corresponding to a field with negative kinetic energy-known as a spin$%
-1$ \textit{phantom} field. This concept was later developed in various
contexts \cite{Visser}. Since then, a wide range of \textit{phantom} black
hole solutions with distinct physical behaviors have been investigated \cite%
{PBH1,PBH2,PBH3,PBH4,PBH5,PBH6,PBH7,PBH8,PBH9,PBH10,PBH11,PBH12,PBH13,PBH14,PBH15}%
.

The remarkable similarity between the thermodynamic behavior of charged AdS
black holes and that of a liquid-gas system was first observed
by Chamblin \textit{et al.} \cite{Chamblin1,Chamblin2}, who highlighted the
van der Waals-like phase transition structure in such systems.
Subsequently, Kubiznak and Mann \cite{Kubiznak2012} extended this analogy by interpreting the cosmological
constant $\Lambda$ as the pressure in the thermodynamic phase space, where
the conjugate pair of volume and pressure forms a natural thermodynamic
pair. This formulation gave rise to extensive studies of the $P-V$
criticality in black hole thermodynamics.

The phase transition of black holes can be studied through two complementary
approaches: the classical thermodynamic method and the geometrical
thermodynamic framework. In the classical perspective, similarities between
AdS black hole phase transitions and van der Waals fluids have been
established through analogies with the Ehrenfest equations. The pioneering
work of Banerjee \textit{et al.} \cite%
{Banerjee2011A,Banerjee2011B,Banerjee2012} generalized Ehrenfest relations
for black holes, identifying key conjugate variables such as $V
\leftrightarrow Q$ and $P \leftrightarrow -U$, thereby enabling black hole
systems to be viewed within a grand-canonical ensemble and their phase
transitions to be analyzed accordingly.

The motion of test particles in any curved spacetime follows the geodesic
equations, which encode the underlying structure of the geometry. These
equations are typically nonlinear and challenging to solve analytically \cite%
{Weinberg,Wald}. However, in certain special backgrounds---particularly near
compact objects---the geodesic paths can be expressed in terms of elliptic
and hyperelliptic functions. Historically, this approach was first
introduced by Hagihara (1931) \cite{Hagihara}, who represented the
particle's trajectory in the Schwarzschild geometry using elliptic
functions, later extended by Jacobi and Weierstrass \cite{Jacobi,Weierstrass}%
. In addition, Nojiri and Odintsov in an intriguing study \cite%
{ShadowF(R)} explored the radii of the photon sphere and the black hole
shadow in the context of $F(R)$ gravity, specifically examining general
spherically symmetric and static configurations in four dimensions. In this
research, the scalar curvature $R$ was expressed as a function of the radial
coordinate $r$. The study also discussed the implications of these findings
for understanding the black hole shadow. Finally, it identified the
parameter regions that are consistent with observations of M$87^{*}$ and Sgr 
$A^{*}$.

Given the theoretical importance of low-dimensional black holes and the role
of nonlinear electrodynamics and $F(R)$ gravity, the present work aims to
derive exact \textit{phantom} BTZ black hole solutions within the combined 
\textit{$F(R)-$conformal invariant Maxwell} framework. We further examine
how the \textit{phantom} field affects fundamental physical characteristics
of the model, including the event horizon structure, conserved and
thermodynamic quantities, along with overall thermal stability. Subsequent
sections of this paper provide detailed analyses of Ehrenfest equations for
these solutions and thoroughly explore the influence of the \textit{phantom}
field on the geodesic dynamics of test particles near the black hole.

\section{The field equations and black hole solutions}

Below, we present the governing action that encapsulates the coupling
between $F(R)$ gravity and the \textit{power-Maxwell} field when formulated
in a three-dimensional spacetime 
\begin{equation}
\mathcal{I}=\int_{\partial \mathcal{M}}d^{3}x\sqrt{-g}\left[ F(R)-2\kappa
^{2}\eta \left( -\mathcal{F}\right) ^{s}\right] ,  \label{actionF(R)}
\end{equation}%
the initial component of the aforementioned action corresponds to the $F(R)$
gravitational framework, formally structured as $F(R)=R+f(R)$. In this
structure, $R$ signifies the scalar curvature, and $f(R)$ is designated as
an unconstrained function of this curvature. The subsequent term details the
interaction: it models a coupling with the \textit{power-Maxwell} field when
the parameter $\eta$ is set to $+1$, or alternatively, with a spin$-1$ 
\textit{phantom} field if $\eta$ equals $-1$. The exponent $s$ in the 
\textit{power-Maxwell} formalism is represented by $s$. Crucially, $\mathcal{%
F}=F_{\mu \nu}F^{\mu \nu}$ stands as the defining Maxwell invariant.
Furthermore, the electromagnetic tensor field ($F_{\mu \nu }$) is defined
via the standard gauge expression $F_{\mu \nu}=\partial _{\mu }A_{\nu
}-\partial _{\nu }A_{\mu }$, where $A_{\mu }$ is the associated gauge
potential. The coupling constant is $\kappa ^{2}=8\pi G$, with $G$
representing the gravitational constant of Newton. Within the scope of this
action, $g=\det (g_{\mu \nu })$ denotes the determinant of the metric tensor 
$g_{\mu \nu }$. For all subsequent analysis, we utilize the unified units
where $G=c=1$.

The dynamical field equations governing $F(R)$ gravity are obtained by
applying the principle of minimal action, achieved through independent
variation of the action functional, $\mathcal{I}$, in relation to both the
metric tensor $g_{\mu\nu }$ and the $U(1)$ gauge potential $A_{\mu }$. This
variational procedure yields the set of governing equations 
\begin{eqnarray}
R_{\mu \nu }\left( 1+f_{R}\right) -\frac{g_{\mu \nu }F(R)}{2}+\left( g_{\mu
\nu }\nabla ^{2}-\nabla _{\mu }\nabla _{\nu }\right) f_{R} &=&8\pi T_{\mu
\nu },  \label{EqF(R)1} \\
&&  \notag \\
\partial _{\mu }\left( \sqrt{-g}\left( -\mathcal{F}\right) ^{s-1}F^{\mu \nu
}\right) &=&0,  \label{EqF(R)2}
\end{eqnarray}%
where $f_{R}=\frac{df(R)}{dR}$. In Eq. (\ref{EqF(R)1}), $T_{\mu \nu }$ is
related to the energy--momentum tensor of \textit{power-Maxwell} field,
where is defined as 
\begin{equation}
T_{\mu \nu }=\frac{-\eta }{4\pi }\left( s\left( -\mathcal{F}\right)
^{s-1}F_{\mu }^{~\alpha }F_{\nu \alpha }+\frac{1}{4}g_{\mu \nu }\left( -%
\mathcal{F}\right) ^{s}\right) .
\end{equation}

Whereas we are interested to obtain the \textit{phantom} BTZ black holes, so
we consider a three-dimensional spacetime, represented as 
\begin{equation}
ds^{2}=-\psi (r)dt^{2}+\frac{dr^{2}}{\psi (r)}+r^{2}d{\varphi }^{2},
\label{Metric}
\end{equation}%
in which $\psi (r)$ is referred to as the metric function. Achieving a
precise analytical resolution for $F(R)$ gravity in the presence of coupling
to a matter field demands a careful selection of the source dynamics. We
mandate that the source must manifest as a \textit{conformally invariant}
(anti-)Maxwell field, also referred to as the \textit{phantom} field. A
direct implication of this source formulation is the conservation property
that renders the energy-momentum tensor entirely traceless. It is important
to note that the \textit{power-Maxwell} field becomes the \textit{%
conformally invariant Maxwell} (or \textit{phantom}) field when $s=\frac{d}{4%
}$, where $d$ is the dimension of spacetime. Specifically, in
three-dimensional spacetime, the \textit{power-Maxwell} field transitions to
the \textit{conformally invariant Maxwell} (or \textit{phantom}) field when $%
s=\frac{3}{4}$. Consequently, to develop the \textit{$F(R)$-conformally
invariant Maxwell} (or \textit{phantom}) theory of gravity, we need to
substitute $s=\frac{3}{4}$ into Eq. (\ref{EqF(R)2}), which leads to

\begin{equation}
\partial _{\mu }\left( \sqrt{-g}\left( -\mathcal{F}\right) ^{-1/4}F^{\mu \nu
}\right) =0,  \label{CIM}
\end{equation}%
where $T_{\mu \nu }$ in Eq. (\ref{EqF(R)1}) is given by 
\begin{equation}
T_{\mu \nu }=\frac{-\eta }{4\pi }\left( \frac{3}{4}\left( -\mathcal{F}%
\right) ^{-1/4}F_{\mu }^{~~\alpha }F_{\nu \alpha }+\frac{1}{4}g_{\mu \nu
}\left( -\mathcal{F}\right) ^{3/4}\right) .  \label{CIMphantom}
\end{equation}

In addition, our objective is to derive the exact solutions under the
constraint of a constant scalar curvature, $R=R_{0}$, specifically within
the context of three-dimensional $F(R)$ gravity coupled with the \textit{%
conformally invariant Maxwell} (or \textit{phantom}) field. By taking the
trace of Eq. (\ref{EqF(R)1}), we establish the necessary algebraic
condition: $R_{0}\left( 1+f_{R_{0}}\right) -\frac{3}{2}\left(
R_{0}+f(R_{0})\right) =0$, where $f_{R_{0}}$ is defined as the value of the
derivative $f_{R}$ evaluated at $R=R_{0}$. This equation then determines the
required value for $R_{0}$, which leads to 
\begin{equation}
R_{0}=\frac{3f(R_{0})}{2f_{R_{0}}-1}.  \label{R0}
\end{equation}

The dynamical equations governing the \textit{$F(R)-$conformally invariant
Maxwell} (or \textit{phantom)} theory of gravity are ascertained by
substituting the result from Equation (\ref{R0}) back into Eq. (\ref{EqF(R)1}%
), leading to an alternative, reformulated expression for the governing laws
of motion 
\begin{eqnarray}
&&R_{\mu \nu }\left( 1+f_{R_{0}}\right) -\frac{g_{\mu \nu }}{3}R_{0}\left(
1+f_{R_{0}}\right) =-2\eta \left( \frac{3}{4}\left( -\mathcal{F}\right)
^{-1/4}F_{\mu }^{~~\alpha }F_{\nu \alpha }+\frac{1}{4}g_{\mu \nu }\left( -%
\mathcal{F}\right) ^{3/4}\right) .  \label{F(R)Trace}
\end{eqnarray}

To characterize the stationary solutions corresponding to electrically
charged black holes, we introduce a radial electric field characterized by
the gauge potential $A_{\mu }=h\left( r\right) \delta _{\mu }^{t}$. The
coupling of this ansatz with the field equations (Eqs. (\ref{CIM}) and (\ref%
{Metric})) results in the differential constraint: $rh^{\prime \prime
}(r)+2h^{\prime }(r)=0$ (where derivatives are taken with respect to the
radial coordinate $r$). The integration of this equation provides the
specific radial dependence $h(r)=-\frac{q^{2/3}}{r}$, where $q$ serves as
the integration constant linked to the total electric charge. Utilizing this
established $h(r)$, the formal expression for the electromagnetic field
tensor is then constructed as 
\begin{equation}
F_{\mu \nu }=\left( 
\begin{array}{ccc}
0 & \frac{q^{2/3}}{r^{2}} & 0 \\ 
-\frac{q^{2/3}}{r^{2}} & 0 & 0 \\ 
0 & 0 & 0%
\end{array}%
\right) .  \label{E(r)}
\end{equation}

By synthesizing the information contained within the introduced metric (\ref%
{Metric}), the constraint provided by the trace of the field equations (\ref%
{F(R)Trace}), and the expression for the electromagnetic field tensor (\ref%
{E(r)}) , we seek the exact solutions characterizing the metric function $%
\psi (r)$. Following the necessary calculations, the mathematical
relationship governing $\psi (r)$ manifests itself as the following set of
coupled differential equations 
\begin{eqnarray}
eq_{tt} &=&eq_{rr}=r^{2}\left( 1+f_{R_{0}}\right) \left( \frac{2rR_{0}}{3}%
+r\psi ^{\prime \prime }(r)+\psi ^{\prime }(r)\right)+\frac{\eta q}{2^{1/4}},
\label{eq11} \\
&&  \notag \\
eq_{\varphi \varphi } &=&r\left( 1+f_{R_{0}}\right) \left( \psi ^{\prime
}(r)+\frac{rR_{0}}{3}\right) -\frac{\eta q}{2^{1/4}},  \label{eq22}
\end{eqnarray}%
We proceed by substituting the definitions of $eq_{tt}$, $eq_{rr}$, and $%
eq_{\varphi \varphi }$ (the respective $tt$, $rr$, and $\varphi \varphi $
parts of field equation (\ref{F(R)Trace})) into the framework where the
scalar curvature is held constant ($R=R_{0}$). This specific setup enables
us to derive the following form for the metric function through the
application of Eqs. (\ref{eq11}) and (\ref{eq22}) 
\begin{equation}
\psi (r)=-m_{0}-\frac{R_{0}r^{2}}{6}-\frac{\eta q}{2^{1/4}\left(
1+f_{R_{0}}\right) r},  \label{g(r)F(R)}
\end{equation}%
where all tensor components dictated by the field equations (\ref{F(R)Trace}%
) are successfully satisfied by the solution established in (\ref{g(r)F(R)}%
). This solution inherently incorporates the term $m_{0}$, which serves as
the integration constant representing the total mass of the black hole. A
necessary prerequisite for obtaining physically meaningful results is the
constraint $f_{R}\neq -1$.

To ascertain the existence of a singularity, one key analytical quantity is
the Kretschmann scalar. By employing the three-dimensional spacetime defined
via Eq. (\ref{Metric}) and the corresponding metric function (\ref{g(r)F(R)}%
), the Kretschmann scalar is obtained in the configuration 
\begin{equation}
R_{\alpha \beta \gamma \delta }R^{\alpha \beta \gamma \delta }=\frac{%
R_{0}^{2}}{3}+\frac{3\sqrt{2}\eta ^{2}q^{2}}{\left( 1+f_{R_{0}}\right)
^{2}r^{6}},
\end{equation}%
where reveals that the Kretschmann scalar becomes singular at $r=0$, a
direct consequence of the electrical interaction term (the second term).
Mathematically, this is encapsulated by the limit $\lim_{r\longrightarrow
0}R_{\alpha \beta \gamma \delta }R^{\alpha \beta \gamma
\delta}\longrightarrow \infty $, a definitive indicator of a curvature
singularity situated at the origin. The scalar quantity is well-defined and
finite for any non-zero radial coordinate. In addition, its behavior in the
asymptotic limit is provided by 
\begin{equation}
\lim_{r\longrightarrow \infty }R_{\alpha \beta \gamma \delta }R^{\alpha
\beta \gamma \delta }\longrightarrow \frac{R_{0}^{2}}{3}.
\end{equation}
Furthermore, the asymptotic trend of the metric function dictates that 
\begin{equation}
\lim_{r\longrightarrow \infty }\psi \left( r\right) \longrightarrow \frac{%
-R_{0}r^{2}}{6},
\end{equation}%
and by setting $R_{0}=6\Lambda$, it is shown that the spacetime structure
approaches an asymptotically AdS form.

\begin{figure}[tbph]
\centering
\includegraphics[width=0.45\linewidth]{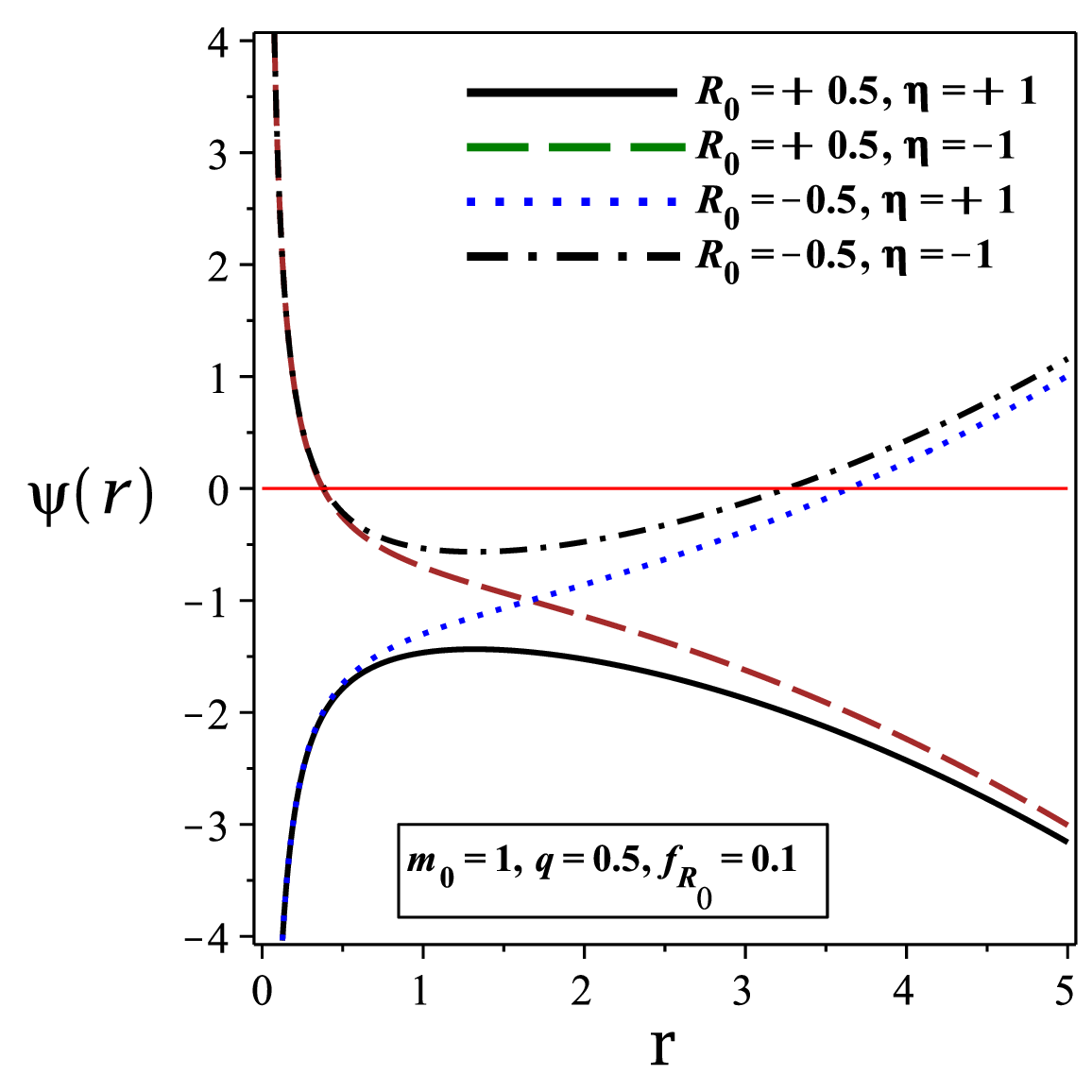} \newline
\caption{The metric function $\protect\psi (r)$ versus $r$ for Maxwell ($%
\protect\eta=+1$), and \textit{phantom} ($\protect\eta=-1$) fields.}
\label{Fig1}
\end{figure}

To investigate the effects of $R_{0}$ and $\eta $ on the obtained solution,
we plot the metric function ($\psi (r)$) versus $r$ in Figure. \ref{Fig1}.
Our analysis shows that the solutions obtained in Eq. (\ref{g(r)F(R)}), can
include an event horizon when $R_{0}$ is negative. This leads to two
distinct behaviors for BTZ black holes depending on whether $R_{0}$ is
negative or positive:

1- \textbf{\emph{For $R_{0}>0$:}} The solutions yield a real root that does
not correspond to the event horizon. Therefore, for $R_{0}>0$, no BTZ black
hole solutions exist (as indicated by the continuous and dashed lines in
Fig. \ref{Fig1}). Notably, references \cite{BTZNON1, BTZNON2, BTZNON3}
demonstrate that BTZ dS-black holes cannot exist. Our findings extend this
conclusion, showing that Maxwell and phantom (anti-Maxwell) BTZ black holes
in $F(R)$ gravity also cannot exist when $R_{0}>0$.

2- \textbf{\emph{For $R_{0}<0$:}} It is possible for the singularity to be
enveloped by an event horizon. Specifically, Maxwell and \textit{phantom}
(anti-Maxwell) BTZ black holes in $F(R)$ gravity do exist when $R_{0}< 0$.
Moreover, there are two distinct behaviors between the Maxwell and \textit{%
phantom} cases (see the dotted and dotted-dashed lines in Fig. \ref{Fig1}).
These distinct behaviors are: i) Maxwell BTZ black holes in $F(R)$ gravity
have only one real root, which corresponds to the event horizon. In
contrast, \textit{phantom} BTZ black holes feature two real positive roots:
the larger root matches the event horizon, and the smaller root is linked to
the inner horizon. ii) For the same parameter values, Maxwell BTZ black
holes (represented by the dotted line in Fig. \ref{Fig1}) are larger than 
\textit{phantom} BTZ black holes (depicted by the dotted-dashed line in Fig. %
\ref{Fig1}).

%%%%%%%%%%%%%%%%%%%%%%%%%%%%%%%%%%%%%%%%%%%%%%%%%%%%%%%%%%%%%%%
%%%%%%%%%%%%%%%%%%%%%%%%%%%%%%%%%%%%%%%%%%%%%%%%%%%%%%%%%%%%%%%

\section{Thermodynamics}

Expressing the mass ($m_{0}$) in terms of the event horizon radius ($r_{+}$%
), scalar curvature ($R_{0}$), electric charge ($q$), and the $F(R)$ gravity
parameter (as presented below) is essential for analyzing the thermodynamic
properties of the obtained black hole solutions. For this purpose, we solve $%
g_{tt}=\psi (r)=0$ to find $m_{0}$, which leads to 
\begin{equation}
m_{0}=-\frac{R_{0}r_{+}^{2}}{6}-\frac{\eta q}{2^{1/4}\left(
1+f_{R_{0}}\right) r_{+}}.  \label{mm}
\end{equation}

First, we determine the surface gravity on the event horizon to subsequently
extract the Hawking temperature ($T$) associated with the \textit{phantom}
BTZ black holes. This critical quantity is provided by 
\begin{eqnarray}
\kappa &=&\left. \frac{g_{tt}^{\prime }}{2\sqrt{-g_{tt}g_{rr}}}=\right\vert
_{r=r_{+}}=\left. \frac{\psi ^{\prime }(r)}{2}\right\vert _{r=r_{+}} =-\frac{%
R_{0}r_{+}}{6}+\frac{\eta q}{2^{5/4}\left( 1+f_{R_{0}}\right) r_{+}^{2}},
\label{k}
\end{eqnarray}%
since $r_{+}$ defines the event horizon radius, the Hawking temperature,
given by $T=\frac{\kappa }{2\pi }$, can then be expressed in the form 
\begin{equation}
T=-\frac{R_{0}r_{+}}{12\pi }+\frac{\eta q}{2^{9/4}\pi \left(
1+f_{R_{0}}\right) r_{+}^{2}},  \label{TemF(R)CPMI}
\end{equation}%
where indicates that $T$ depends on radius of the event horizon, the scalar
curvature, the electric charge, $\eta$ and the parameter of $F(R)$ gravity.

The electric charge of such black holes can be determined through the
application of Gauss's law, which yields the following
result 
\begin{equation}
Q=\frac{3q^{1/3}}{2^{13/4}}.  \label{Q}
\end{equation}

Given that $F_{\mu \nu }=\partial _{\mu }A_{\nu }-\partial _{\nu }A_{\mu }$,
the only non-vanishing component of the gauge potential can be expressed as $%
A_{t}=-\int F_{tr}dr$. Consequently, the electric potential at the
corresponding location is obtained as 
\begin{equation}
U=-\int_{r_{+}}^{+\infty }F_{tr}dr=-\frac{q^{2/3}}{r_{+}}.
\label{elcpoF(R)CPMI}
\end{equation}

To derive the entropy of black holes within the framework of $F(R)$ theory,
one can utilize a modified version of the area law, referred to as the
Noether charge method \cite{F(R)3} 
\begin{equation}
S=\frac{A(1+f_{R_{0}})}{4},  \label{SFR}
\end{equation}%
where $A$ represents the horizon area. In three-dimensional spacetime, the
horizon area is defined as 
\begin{equation}
A=\left. \int_{0}^{2\pi }\sqrt{g_{\varphi \varphi }}\right\vert_{r=r_{+}}=%
\left. 2\pi r\right\vert _{r=r_{+}}=2\pi r_{+}.
\end{equation}
Thus, the entropy of Maxwell--- or \textit{phantom}---BTZ black holes within
the framework of $F(R)$ gravity is given by 
\begin{equation}
S=\frac{\pi (1+f_{R_{0}})r_{+}}{2},  \label{S}
\end{equation}%
this result implies that the area law is not valid for black hole solutions
within the framework of $F(R)$ gravity.

The total mass of these black holes in $F(R)$ gravity is determined via the
Ashtekar-Magnon-Das (AMD) method \cite{AMDI,AMDII}, and its expression is 
\begin{equation}
M=\frac{m_{0}\left( 1+f_{R_{0}}\right) }{8},  \label{AMDMass}
\end{equation}%
substituting the mass from equation (\ref{mm}) into equation (\ref{AMDMass})
results in 
\begin{equation}
M=-\frac{\left( 1+f_{R_{0}}\right) R_{0}r_{+}^{2}}{48}-\frac{\eta q}{%
2^{13/4}r_{+}}.  \label{MM}
\end{equation}
where depends on various parameters, including scalar curvature, the
parameter of $F(R)$ gravity, electric charge, and $\eta$. To explore the
effects of these parameters, we present our findings in Fig. \ref{Fig3}. Our
analysis reveals four distinct behaviors:

i) For Maxwell BTZ black holes, the total mass is consistently negative when 
$R_{0}>0 $ (indicated by the continuous line in Fig. \ref{Fig3}).

ii) For \textit{phantom} BTZ black holes in $F(R)$ gravity, there exists a
root beyond which the total mass becomes negative (illustrated by the dashed
line in Fig. \ref{Fig3}). This implies that large black holes cannot exist
in the \textit{phantom} case when $R_{0}>0$.

iii) Large Maxwell BTZ black holes possess a positive total mass (shown by
the dotted line in Fig. \ref{Fig3}), indicating that they can be considered
physical objects within $F(R)$ gravity.

iv) Consistently positive, the total mass of phantom BTZ black holes within $%
F(R)$ gravity ensures that these objects remain physical, regardless of the
event horizon radius (see the dotted-dashed line in Fig. \ref{Fig3}).

%%%%%%%%%%%%%%%%%%%%%%%%%%%%%%%%%%%%%%%%%%%%%%%%%%%%%%%%%%%%%%%
\begin{figure}[tbph]
\centering
\includegraphics[width=0.45\linewidth]{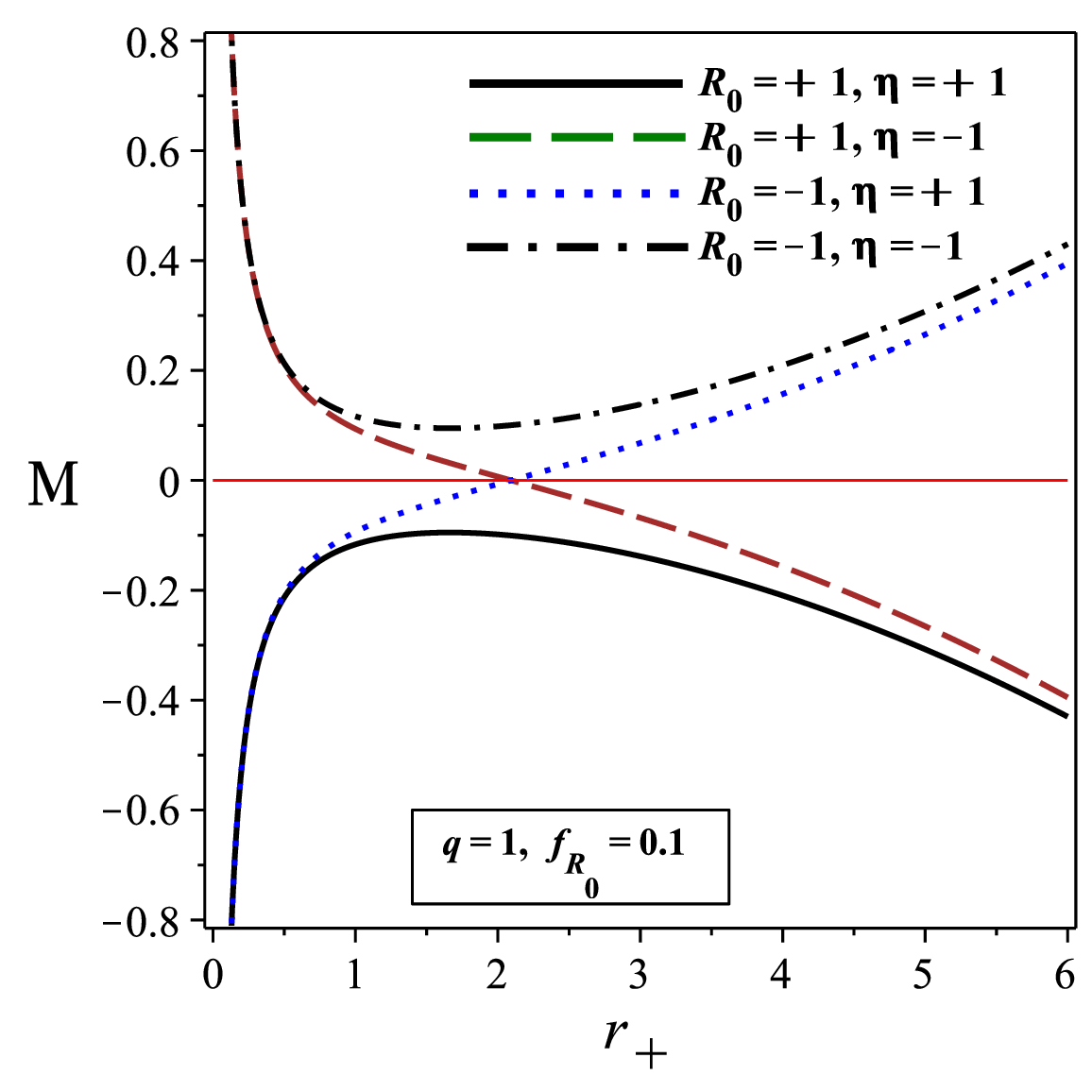} \newline
\caption{The total mass $M$ versus $r_{+}$ for Maxwell ($\protect\eta =+1$),
and \textit{phantom} ($\protect\eta =-1$) cases.}
\label{Fig2}
\end{figure}
%%%%%%%%%%%%%%%%%%%%%%%%%%%%%%%%%%%%%%%%%%%%%%%%%%%%%%%%%%%%%%%

The first law of thermodynamics, $dM=TdS+\eta UdQ$ (with $T=\left( \frac{%
\partial M}{\partial S}\right) _{Q}$ and $\eta U=\left(\frac{\partial M}{%
\partial Q}\right) _{S}$), is straightforwardly satisfied by the conserved
and thermodynamic quantities. These results align precisely with the
calculations presented in Eqs. (\ref{TemF(R)CPMI}) and (\ref{elcpoF(R)CPMI})

\section{Thermal Stability}

We set out to explore the local and global stability of the black hole by
analyzing it through a thermodynamic lens. Later sections detail the impact
that a fixed scalar curvature ($R_{0}$) and the parameter $\eta$ exert on
the local and global stability of \textit{phantom} BTZ solutions in $F(R)$
gravity.

\subsection{Local Stability}

Our primary focus is to explore the local stability of \textit{phantom} BTZ
black holes under the umbrella of $F(R)$ gravity. To achieve this, a
thorough analysis of their heat capacity is essential.

Within the context of the canonical ensemble, it is established that a
thermodynamic system's local stability can be inferred from its heat
capacity. Therefore, we intend to calculate this specific heat capacity for
our solutions and leverage it to evaluate the local stability of the black
holes.

Prior to embarking on the heat capacity calculations, we will first
reformulate the black hole's total mass (Eq. (\ref{MM})) as a function of
its entropy (Eq. (\ref{S})) according to the following expression 
\begin{equation}
M\left( S,Q\right) =\frac{-128\sqrt{2}\pi ^{3}\eta Q^{3}\left(
1+f_{R_{0}}\right) ^{2}-9S^{3}R_{0}}{108\pi ^{2}S\left( 1+f_{R_{0}}\right) },
\label{MSQ}
\end{equation}%
by employing equation (\ref{MSQ}), the temperature expression can be recast
in the subsequent form 
\begin{equation}
T=\left( \frac{\partial M\left( S,Q\right) }{\partial S}\right) _{Q}=\frac{64%
\sqrt{2}\pi ^{3}\eta Q^{3}\left( 1+f_{R_{0}}\right) ^{2}-9S^{3}R_{0}}{154\pi
^{2}S^{2}\left( 1+f_{R_{0}}\right) }.  \label{TM}
\end{equation}

The resulting form for the heat capacity is presented below 
\begin{equation}
C_{Q}=\frac{T}{\left( \frac{\partial T}{\partial S}\right) _{Q}}=\frac{-64%
\sqrt{2}\pi ^{3}\eta Q^{3}\left( 1+f_{R_{0}}\right) ^{2}+9S^{3}R_{0}}{128%
\sqrt{2}\pi ^{3}\eta Q^{3}\left( 1+f_{R_{0}}\right) ^{2}+9S^{3}R_{0}},
\label{Heat1}
\end{equation}

Within the theoretical framework of black hole analysis, a critical
threshold is posited: the heat capacity evaluated precisely at the
zero-temperature limit (i.e., $C_{Q}=T=0$) acts as the demarcation line
separating physically realizable black hole solutions ($T>0$) from those
considered non-physical ($T<0$). This threshold represents a fundamental
constraint on the system's thermodynamic behavior, marked by an inversion in
the sign of the heat capacity. Moreover, the abrupt singularities observed
in the heat capacity function are interpreted as diagnostic markers for
phase transition events within the black hole structure.

The determination of the physical limitation point follows the resolution of
entropy from Eq. (\ref{TM}), presented below 
\begin{equation}
S_{root_{T}}=S_{root_{C_{Q}}}=\frac{4\pi Q}{3R_{0}}\left( 3\sqrt{2}\eta
\left( 1+f_{R_{0}}\right) ^{2}R_{0}^{2}\right) ^{1/3}.
\end{equation}

Achieving the real, positive root necessitates the fulfillment of the
subsequent pair of criteria 
\begin{equation}
S_{root_{T}}>0\rightarrow \left\{ 
\begin{array}{ccc}
R_{0}>0~~\&~~\eta >0 &  & \text{condition I} \\ 
&  &  \\ 
R_{0}<0~~\&~~\eta <0 &  & \text{condition II}%
\end{array}%
\right. .  \label{Sroot2}
\end{equation}

We examine how the parameters $R_{0}$ and $\eta $ affect the roots of
temperature (as shown in Eq. (\ref{TM})). The analysis confirms the
uniqueness of the physical limitation point, which occurs exclusively when $%
\eta R_{0}>0$.

Determining the phase transition criticalities, synonymous with the
divergences in the heat capacity, mandates the solution of the governing
relation $\left( \frac{\partial^{2}M\left( S,Q\right) }{\partial S^{2}}%
\right) _{Q}=0$. So, we get one phase transition critical point in the
following form 
\begin{equation}
S_{div}=\frac{4\pi Q}{3R_{0}}\left( -6\sqrt{2}\eta \left( 1+f_{R_{0}}\right)
^{2}R_{0}^{2}\right) ^{1/3},  \label{rdivHeat}
\end{equation}%
where indicate that for having the real positive divergent point, we have to
respect two following conditions 
\begin{equation}
S_{div_{C_{Q}}}>0\rightarrow \left\{ 
\begin{array}{ccc}
R_{0}>0~~\&~~\eta <0 &  & \text{condition I} \\ 
&  &  \\ 
R_{0}<0~~\&~~\eta >0 &  & \text{condition II}%
\end{array}%
\right. .  \label{Sdiv2}
\end{equation}

\begin{figure}[tbph]
\centering
\includegraphics[width=0.45\linewidth]{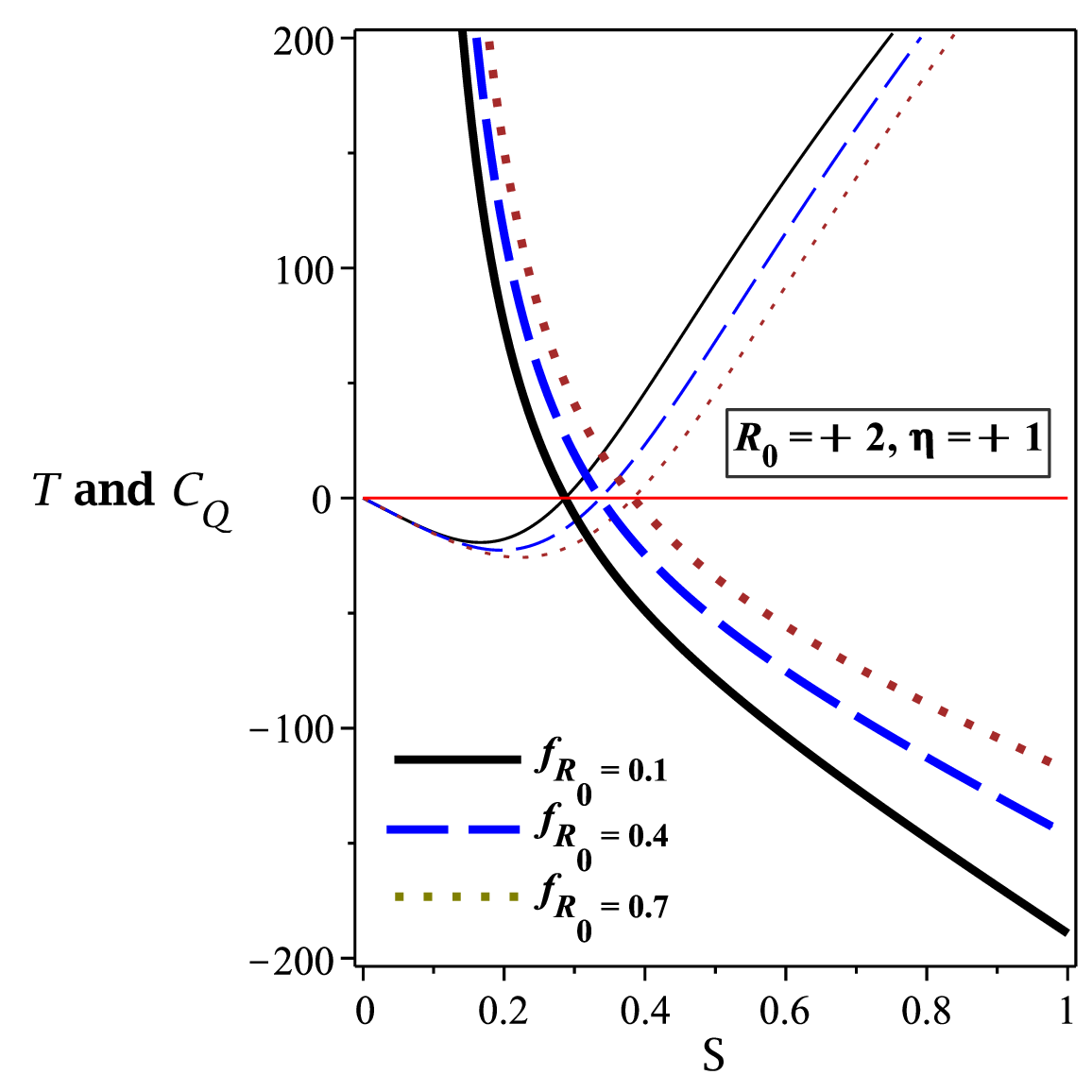} \includegraphics[width=0.45%
\linewidth]{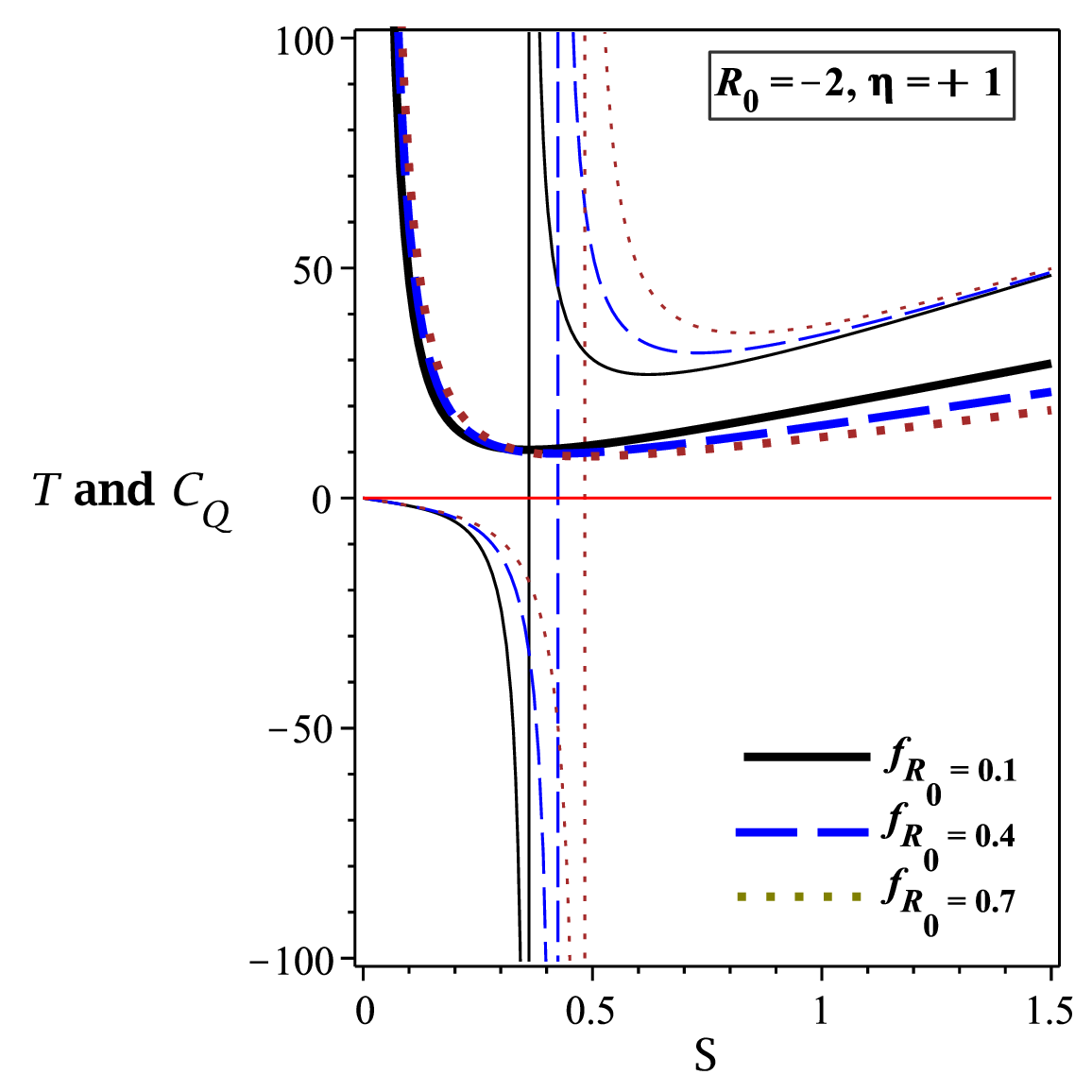}\newline
\includegraphics[width=0.45\linewidth]{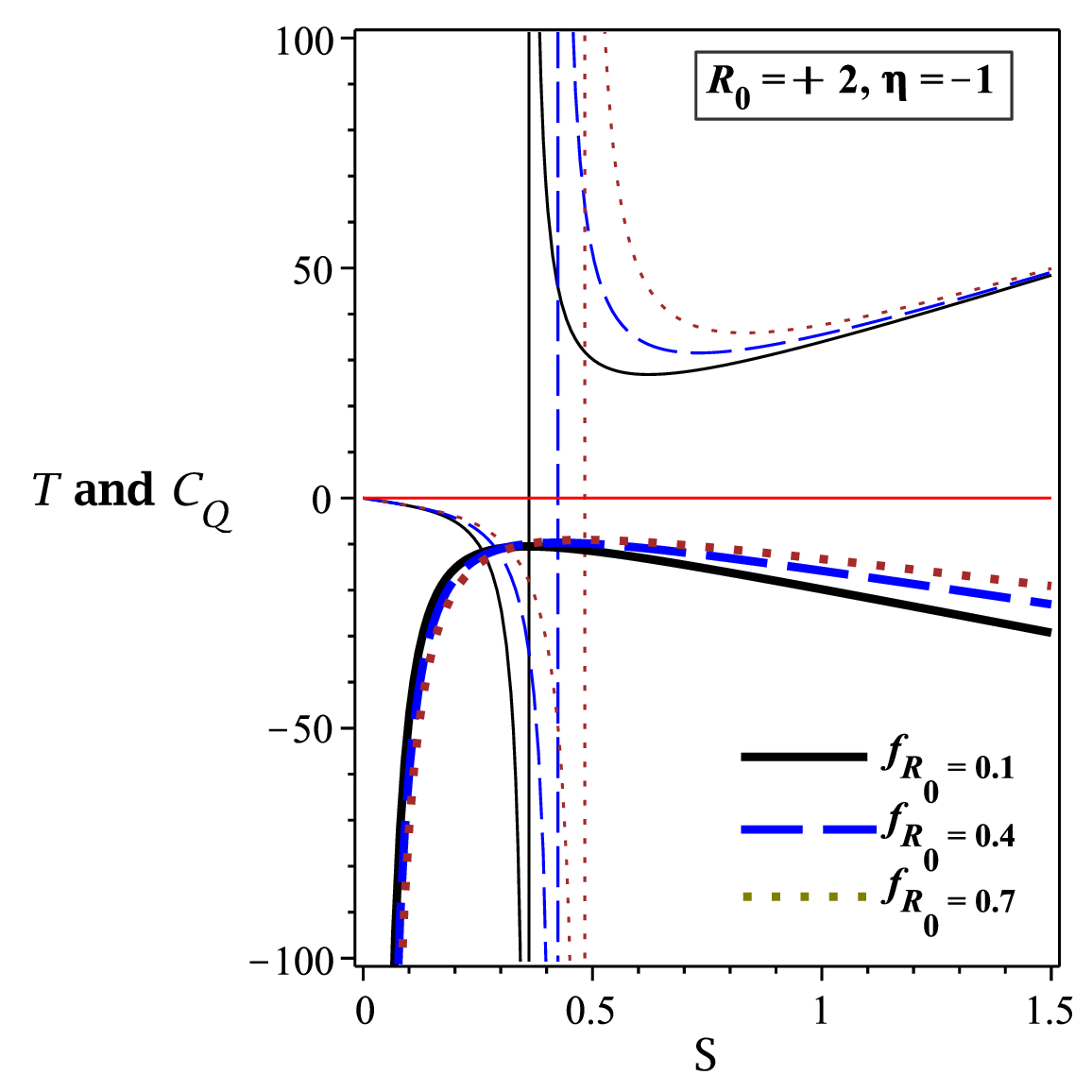} \includegraphics[width=0.45%
\linewidth]{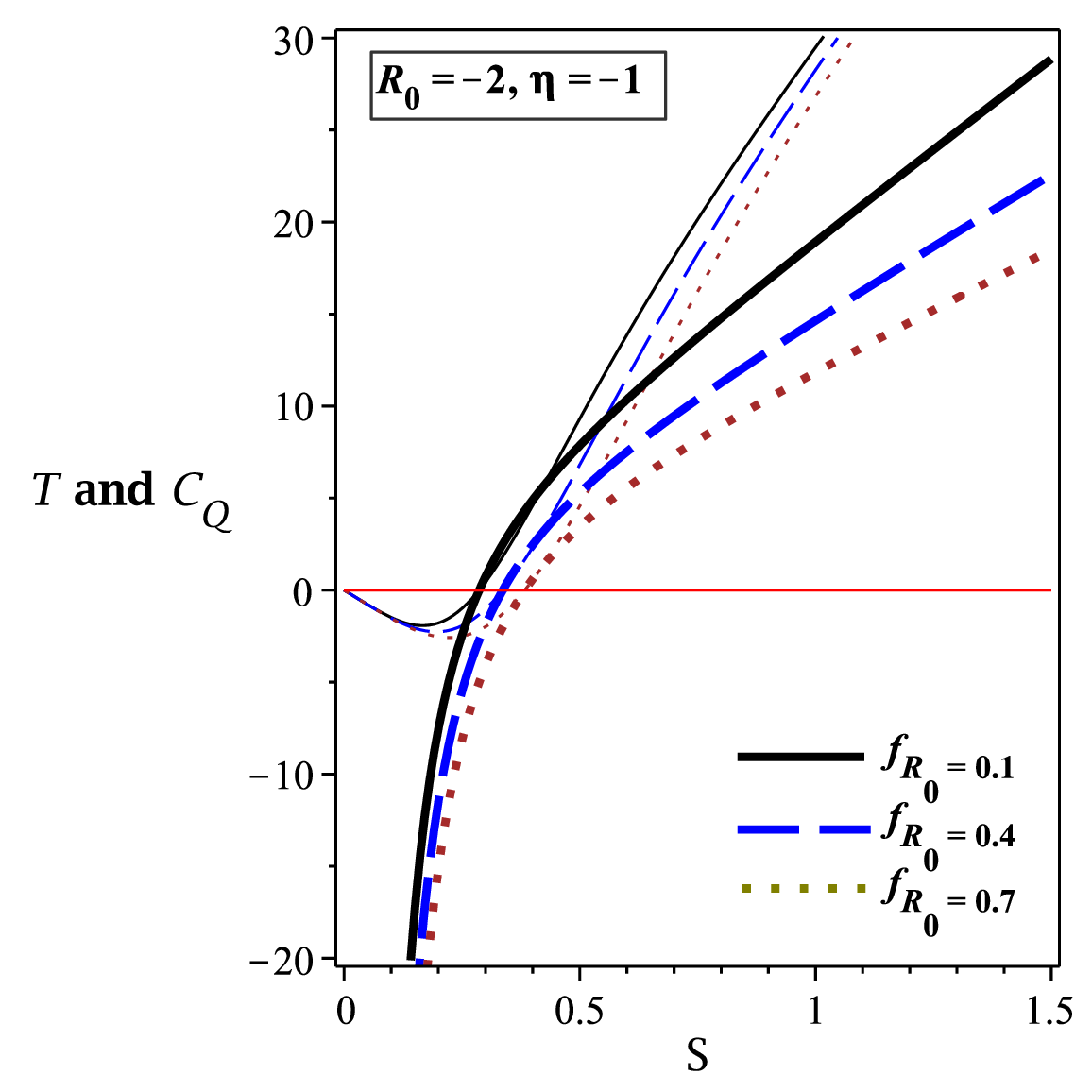}\newline
\caption{The heat capacity $C_{Q}$ (thin lines) and temperature $T$ (thick
lines) versus $S$. Up panels for Maxwell case ($\protect\eta =+1$), and down
panels for \textit{phantom} (anti-Maxwell) case ($\protect\eta =-1$).}
\label{Fig3}
\end{figure}

Our analysis in Eq. (\ref{Sroot2}) shows that \textit{phantom} BTZ black
holes face a physical limitation when $\eta R_{0}>0$. In contrast, these
black holes exhibit a phase transition critical point when $\eta R_{0}<0$
(refer to the conditions outlined in Eq. (\ref{Sdiv2})). Consequently, it is
impossible for both a physical limitation and a phase transition critical
point to occur simultaneously. For further details, we present the
temperature and heat capacity in Fig. \ref{Fig3}.

%%%%%%%%%%%%%%%%%%%%%%%%%%%%%%%%%%%%%%%%%%%%%%%%%%%%%%%%%%%%%%%
%%%%%%%%%%%%%%%%%%%%%%%%%%%%%%%%%%%%%%%%%%%%%%%%%%%%%%%%%%%%%%%

Our findings reveal several intriguing behaviors, including:

1) \textbf{\emph{For $R_{0}>0$}:} Maxwell ($\eta=+1 $) and \textit{phantom} (%
$\eta=-1$) BTZ black holes cannot simultaneously meet the conditions for
being physical and stable (see the up-left panel for the Maxwell case and
the down-left panel for the \textit{phantom} case in Fig. \ref{Fig3}).
Specifically, both the temperature ($T$) and heat capacity ($C_{Q}$) cannot
be positive at the same time. Furthermore, there are distinct behaviors
between Maxwell and \textit{phantom} BTZ black holes in $F(R)$ gravity,
which are linked to the presence of physical limitations and phase
transition points. In the Maxwell case, a physical limitation point exists,
while the \textit{phantom} black holes experience a phase transition point
between small and large radii. Consequently, Maxwell and \textit{phantom}
BTZ black holes in $F(R)$ gravity cannot exist when $R_{0}$ is positive.

2) \textbf{\emph{For $R_{0}<0$}:} Maxwell ($\eta=+1 $) and \textit{phantom} (%
$\eta=-1$) BTZ black holes, characterized by large entropy or large radius,
are physical and stable objects.

There are two distinct behaviors observed in Maxwell and \textit{phantom}
BTZ black holes in $F(R)$ gravity:

i) For Maxwell BTZ black holes, a phase transition point exists between
small unstable and large stable configurations (see the up-right panel in
Fig. \ref{Fig3}). In contrast, for \textit{phantom} BTZ black holes, there
is a physical limitation point separating small unstable and large stable
configurations (see the down-right panel in Fig. \ref{Fig3}).

ii) A comparison between the heat capacity's phase transition point (for
Maxwell BTZ black holes) and the corresponding physical limitation point
(for \textit{phantom} BTZ black holes) demonstrates that the stable and
physically admissible domain for \textit{phantom} BTZ black holes is more
extensive.

\subsection{Global Stability}

Within the grand-canonical ensemble, the Gibbs potential is the primary tool
for probing global stability, as the condition $G<0$ is prerequisite for its
robustness. Therefore, our strategy for analyzing \textit{phantom} BTZ black
holes in the $F(R)$ background is oriented toward the precise quantification
of this very Gibbs potential

We extract the Gibbs potential as 
\begin{eqnarray}
G &=&M\left( S,Q\right) -TS-\eta UQ  \notag \\
&&  \notag \\
&=&\frac{128\sqrt{2}\pi ^{3}\eta Q^{3}\left( 1+f_{R_{0}}\right)
^{2}+9S^{3}R_{0}}{108\pi ^{2}S\left( 1+f_{R_{0}}\right) },  \label{Gibbs}
\end{eqnarray}%
where we can get the root of the Gibbs potential as 
\begin{equation}
S_{root_{G}}=\frac{4\pi Q}{3R_{0}}\left( -6\sqrt{2}\eta \left(
1+f_{R_{0}}\right) ^{2}R_{0}^{2}\right) ^{1/3},
\end{equation}%
using the equation above and the fact that BTZ black holes exist only for $%
R_{0}<0$, we can identify a real root solely for Maxwell BTZ black holes ($%
\eta =+1$). This real root of the Gibbs potential depends on $R_{0}$, $Q$,
and $f_{R_{0}}$. In contrast, there is no real root of the Gibbs potential
for \textit{phantom} BTZ black holes.

\begin{figure}[tbph]
\centering
\includegraphics[width=0.45\linewidth]{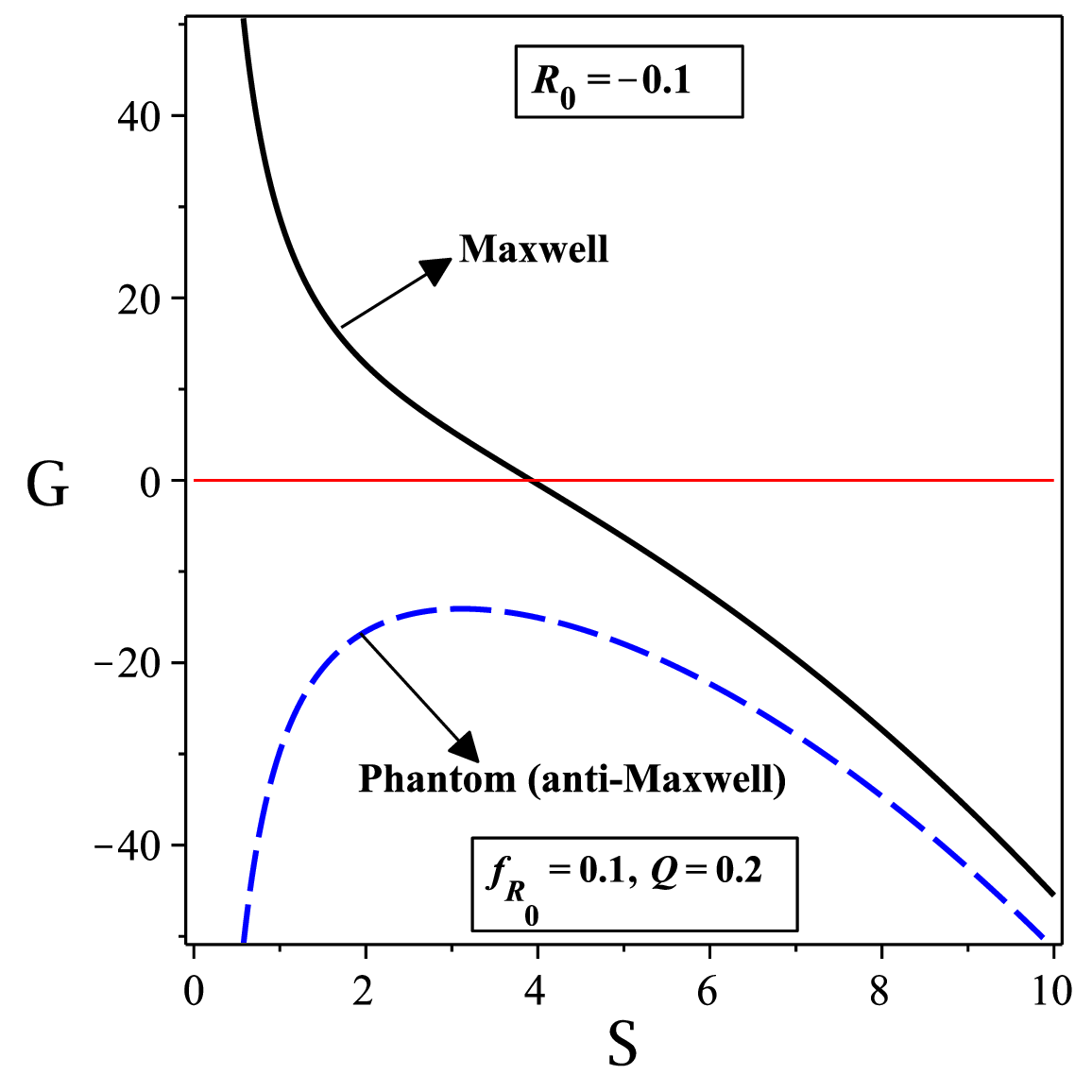}\newline
\caption{The Gibbs potential $G$ (Eq. (\protect\ref{Gibbs})) versus $S$.}
\label{Fig4}
\end{figure}

To study global stability, we must consider the case when $G < 0$. Black
holes exhibit global stability under this condition. To evaluate the global
stability of Maxwell and \textit{phantom} BTZ black holes in $F(R)$ gravity,
we plot the Gibbs potential in Fig. \ref{Fig4}. Our analysis yields the
following results:

i) \textbf{\emph{Maxwell BTZ black holes ($\eta =+1$)}}: There exists a root
for the Gibbs potential that divides the Maxwell BTZ black holes into two
distinct regions. For $S<S_{root_{G}}$, the Gibbs potential is positive,
indicating that Maxwell BTZ black holes cannot satisfy the global stability
condition. In contrast, for $S>S_{root_{G}}$, the Gibbs potential becomes
negative, signifying that Maxwell BTZ black holes are globally stable
objects (see the continuous line in Fig. \ref{Fig4}).

ii) \textbf{\emph{\textit{phantom} BTZ black holes ($\eta =-1$)}}: For these
In this case, there is no root, and the Gibbs potential remains negative at
all points. Thus, \textit{phantom} BTZ black holes consistently satisfy the
global stability condition (see the dashed line in Fig. \ref{Fig4}).

%%%%%%%%%%%%%%%%%%%%%%%%%%%%%%%%%%%%%%%%%%%%%%%%%%%%%%%%%%%%%%%
%%%%%%%%%%%%%%%%%%%%%%%%%%%%%%%%%%%%%%%%%%%%%%%%%%%%%%%%%%%%%%%

\section{Analytical Study of Ehrenfest Relations in Extended Phase Space}

The demarcation between first-order and higher-order phase transitions
within the framework of classical thermodynamics hinges upon the
differential relations utilized: specifically, the Clausius-Clapeyron
equation serves as the necessary and sufficient condition for characterizing
a first-order transformation, whereas second-order transitions are
delineated by adherence to the Ehrenfest relations.

Banerjee \textit{et al.} \cite{Banerjee2012} presented a compelling analogy
of the Ehrenfest equations by comparing thermodynamic state
variables--specifically, $V\leftrightarrow Q$ and $P\leftrightarrow -U$%
--with black hole parameters. This comparison leads to 
\begin{eqnarray}
-\left( \frac{\partial U}{\partial T}\right) _{S} &=&\frac{%
C_{U_{2}}-C_{U_{1}}}{TQ\left( \alpha _{2}-\alpha _{1}\right) }=\frac{\Delta
C_{U}}{TQ\Delta \alpha }, \\
&&  \notag \\
-\left( \frac{\partial U}{\partial T}\right) _{Q} &=&\frac{\alpha
_{2}-\alpha _{1}}{\kappa _{T_{2}}-\kappa _{T_{1}}}=\frac{\Delta \alpha }{%
\Delta \kappa _{T}},
\end{eqnarray}%
where $U$ is the electric potential. Also, $V$ is the thermodynamic volume
of the black hole. In addition, $\alpha $ and $\kappa _{T}$ are the analogs
of the volume expansion coefficient and the isothermal compressibility,
respectively, and are defined as 
\begin{eqnarray}
\alpha &=&\frac{1}{Q}\left( \frac{\partial Q}{\partial T}\right) _{U}, \\
&&  \notag \\
\kappa _{T} &=&\frac{1}{Q}\left( \frac{\partial Q}{\partial U}\right) _{T}.
\end{eqnarray}

Since the $P-V$ criticality of black holes is inherently connected to their
specific heat at constant pressure, $C_{P}$, this correlation renders the
classical Ehrenfest equations directly applicable to the analysis of black
hole phase transitions. Therefore, in this section, we will analytically
verify the following Ehrenfest equations 
\begin{eqnarray}
\left( \frac{\partial P}{\partial T}\right) _{S} &=&\frac{C_{P_{2}}-C_{P_{1}}%
}{VT\left( \alpha _{2}-\alpha _{1}\right) }=\frac{\Delta C_{P}}{VT\Delta
\alpha },  \label{Ehrenfest1} \\
&&  \notag \\
\left( \frac{\partial P}{\partial T}\right) _{V} &=&\frac{\alpha _{2}-\alpha
_{1}}{\kappa_{T_{2}}-\kappa_{T_{1}}}=\frac{\Delta \alpha }{\Delta \kappa_{T}}%
,  \label{Ehrenfest2}
\end{eqnarray}%
within the thermodynamic framework, $\alpha$ represents the coefficient of
volume expansion, while $\kappa_{T}$ signifies isothermal compressibility,
both defined through the expressions given below \cite{Mo2013}. 
\begin{eqnarray}
\alpha &=&\frac{1}{V}\left( \frac{\partial V}{\partial T}\right) _{P},
\label{alpha} \\
&&  \notag \\
\kappa_{T} &=&\frac{-1}{V}\left( \frac{\partial V}{\partial P}\right) _{T}.
\label{kT}
\end{eqnarray}

We will calculate the relevant quantities in equations (\ref{Ehrenfest1})-(%
\ref{Ehrenfest2}) to analyze the $P-V$ criticality in the extended phase
space of \textit{phantom} BTZ black holes.

By replacing $R_{0}=-48\pi P$ within Eq. (\ref{MSQ}), we find that 
\begin{equation}
M\left( S,Q,P\right) =\frac{108S^{3}P-32\sqrt{2}\pi ^{2}\eta Q^{3}\left(
1+f_{R_{0}}\right) ^{2}}{27\pi S\left( 1+f_{R_{0}}\right) }.  \label{MSQP}
\end{equation}

Using Eq. (\ref{MSQP}), we can obtain the temperature ($T$), thermodynamic
volume ($V$), and specific heat of black holes at constant pressure ($C_{P}$%
) in the following forms 
\begin{eqnarray}
T &=&\left( \frac{\partial M\left( S,Q,P\right) }{\partial S}\right) =\frac{%
216S^{3}P+32\sqrt{2}\pi ^{2}\eta Q^{3}\left( 1+f_{R_{0}}\right) ^{2}}{27\pi
S^{2}\left( 1+f_{R_{0}}\right) },  \label{TSPQ} \\
&&  \notag \\
V &=&\left( \frac{\partial M\left( S,Q,P\right) }{\partial P}\right) =\frac{%
4S^{2}}{\pi \left( 1+f_{R_{0}}\right) },  \label{V} \\
&&  \notag \\
C_{P} &=&T\left( \frac{\partial S}{\partial T}\right) _{P}=\frac{S\left(
27S^{3}P+4\sqrt{2}\pi ^{2}\eta Q^{3}\left( 1+f_{R_{0}}\right) ^{2}\right) }{%
27S^{3}P-8\sqrt{2}\pi ^{2}\eta Q^{3}\left( 1+f_{R_{0}}\right) ^{2}}.
\label{CP}
\end{eqnarray}

We can obtain $\alpha$ and $\kappa_{T}$ by applying Eqs. (\ref{TSPQ}) and (%
\ref{V}) in conjunction with Eqs. (\ref{alpha}) and (\ref{kT}). This results
in 
\begin{eqnarray}
\alpha &=&\frac{1}{V}\left( \frac{\partial V}{\partial T}\right) _{P}=\frac{%
27\pi S^{2}\left( 1+f_{R_{0}}\right) }{4\left( 27S^{3}P-8\sqrt{2}\pi
^{2}\eta Q^{3}\left( 1+f_{R_{0}}\right) ^{2}\right) },  \label{alpha2} \\
&&  \notag \\
\kappa_{T} &=&\frac{-1}{V}\left( \frac{\partial V}{\partial P}\right) _{T}=%
\frac{54S^{3}}{27S^{3}P-8\sqrt{2}\pi ^{2}\eta Q^{3}\left( 1+f_{R_{0}}\right)
^{2}}.  \label{kT2}
\end{eqnarray}

It is important to note that $C_{P}$, $\alpha$, and $\kappa_{T}$ have a
shared denominator: $27S^{3}P - 8\sqrt{2}\pi^{2}\eta Q^{3}(1 +
f_{R_{0}})^{2} $. This implies that both $\alpha$ and $\kappa_{T}$ may
diverge at the critical point, similar to how the specific heat at constant
pressure behaves when $27S^{3}P - 8\sqrt{2}\pi^{2}\eta Q^{3}(1 +
f_{R_{0}})^{2} = 0$.

We will now examine the validity of the Ehrenfest equations (Eqs. (\ref%
{Ehrenfest1}) and (\ref{Ehrenfest2})) at the critical point. To do this, we
will utilize the definition of the volume expansion coefficient $\alpha$
(Eq. (\ref{alpha})), leading to the following relationship 
\begin{equation}
V\alpha =\left( \frac{\partial V}{\partial T}\right) _{P}=\left( \frac{%
\partial V}{\partial S}\right) _{P}\left( \frac{\partial S}{\partial T}%
\right) _{P}=\left( \frac{\partial V}{\partial T}\right) _{P}\frac{C_{P}}{T},
\end{equation}%
where $\frac{C_{P}}{T}=\left( \frac{\partial S}{\partial T}\right) _{P}$.
Applying the above equation, we rewrite the R.H.S of Eq. (\ref{Ehrenfest1})
as the following form 
\begin{equation}
\frac{\Delta C_{P}}{VT\Delta \alpha }=\left. \left( \frac{\partial S}{%
\partial V}\right) _{P}\right\vert _{c},  \label{RHS}
\end{equation}%
where the footnote "$c$" is related to the values of physical quantities at
the critical point. Applying Eqs. (\ref{S}), and (\ref{V}), within (\ref{RHS}%
), we find that 
\begin{equation}
\frac{\Delta C_{P}}{VT\Delta \alpha }=\frac{\pi \left( 1+f_{R_{0}}\right) }{%
8S_{c}}.  \label{RHS2}
\end{equation}

Using Eq. (\ref{TSPQ}), the L.H.S of Eq. (\ref{Ehrenfest1}) is given by 
\begin{equation}
\left( \frac{\partial P}{\partial T}\right) _{S}=\frac{\pi \left(
1+f_{R_{0}}\right) }{8S_{c}}.  \label{LHS2}
\end{equation}

Equations (\ref{RHS2}) and (\ref{LHS2}) demonstrate that the first Ehrenfest
equation remains valid at the critical point.

We will verify the validity of the second Ehrenfest equation. To do this, we
utilize the thermodynamic identity $\left( \frac{\partial V}{\partial P}%
\right){T} \left( \frac{\partial P}{\partial T}\right){V} \left( \frac{%
\partial T}{\partial V}\right)_{P} = -1$, as reported in Ref. \cite{Mo2013}.
By applying this identity alongside Eqs. (\ref{alpha})-(\ref{kT2}), we can
express the R.H.S of Eq. (\ref{Ehrenfest2}) in the following form 
\begin{equation}
\frac{\Delta \alpha }{\Delta \kappa_{T}}=\left. \left( \frac{\partial P}{%
\partial T}\right) _{P}\right\vert _{c}=\frac{\pi \left( 1+f_{R_{0}}\right) 
}{8S_{c}}.  \label{RHS3}
\end{equation}

The L.H.S of Eq. (\ref{Ehrenfest2}), is given by 
\begin{equation}
\left. \left( \frac{\partial P}{\partial T}\right) _{V}\right\vert _{c}=%
\frac{\pi \left( 1+f_{R_{0}}\right) }{8S_{c}},  \label{LHS3}
\end{equation}%
where we consider the equation (\ref{TSPQ}) to obtain the above relation. By
comparing Eqs. (\ref{RHS3}) and (\ref{LHS3}), we find that the second
Ehrenfest equation valids at the critical point.

The Prigogine--Defay (PD) ratio is defined as \cite{PD1,PD2}%
\begin{equation}
\Pi =\frac{\Delta C_{P}\Delta \kappa _{T}}{VT\left( \Delta \alpha \right)
^{2}},  \label{PD}
\end{equation}%
by replacing Eqs. (\ref{RHS2}) and (\ref{RHS3}) within PD ratio (Eq. (\ref%
{PD})) , we find that 
\begin{equation}
\Pi =1,  \label{PDfinal}
\end{equation}

The PD ratio serves as a quantitative measure for evaluating deviations from
the second Ehrenfest equation. For a typical second-order phase transition,
this ratio is equal to unity \cite{Banerjee2010}. Building on this analysis,
the conclusions derived from Eq. (\ref{PDfinal}) and the inherent
consistency of the Ehrenfest relations allow us to conclude that the
transition at the $P-V$ critical point in the extended phase space of the 
\textit{phantom} BTZ black hole is fundamentally a second-order transition.

\section{Geodesic Structure}

The analysis of geodesic motion provides essential insights into the
spacetime geometry and the influence of modified gravity parameters on
particle dynamics. In particular, in $F(R)$-\textit{phantom} BTZ black
holes, the interplay between curvature corrections and \textit{phantom}
energy modifies the effective potential, leading to qualitatively new
orbital features compared with the standard BTZ solution.

The geodesic motion of a test particle in the spacetime of the \textit{%
phantom} BTZ black hole within the framework of $F(R)$ gravity can be
derived from the Euler-Lagrange formalism. The corresponding Lagrangian
associated with the metric is given by%
\begin{equation}
2\mathcal{L}=g_{\mu \nu }\dot{x}^{\mu }\dot{x}^{\nu }=-\psi (r)\dot{t}^{2}+%
\frac{1}{\psi (r)}\dot{r}^{2}+r^{2}{\dot{\varphi}}^{2},  \notag
\end{equation}%
where the dot denotes differentiation with respect to the affine parameter $%
\tau $ along the geodesic. The Euler$-$Lagrange equations are written as%
\begin{equation}
\frac{\partial \mathcal{L}}{\partial x^{\mu }}-\frac{d}{ds}\left( \frac{%
\partial \mathcal{L}}{\partial \dot{x}^{\mu }}\right) =0,
\end{equation}%
Since $\mathcal{L}$ does not explicitly depend on $t$ and $\varphi $, two
constants of motion arise:%
\begin{eqnarray}
E &=&\psi (r)\dot{t}\Longrightarrow \dot{t}=\frac{E}{\psi (r)},  \label{39}
\\
&&  \notag \\
L &=&r^{2}{\dot{\varphi}}\Longrightarrow {\dot{\varphi}=}\frac{L}{r^{2}},
\label{310}
\end{eqnarray}%
where $E$ and $L$ denote, respectively, the conserved energy and angular
momentum per unit mass of the particle. Using the normalization condition $%
g_{\mu \nu }\dot{x}^{\mu }\dot{x}^{\nu }=-\epsilon $ together with Eqs. (\ref%
{39}) and (\ref{310}), one obtains the radial equation equation for $\dot{r}$%
:

\begin{equation}
\dot{r}^{2}=E^{2}-\left( \epsilon +\frac{L^{2}}{r^{2}}\right) \left[ -m_{0}-%
\frac{R_{0}r^{2}}{6}-\frac{\eta q}{2^{1/4}\left( 1+f_{R_{0}}\right) r}\right]
,  \label{rad}
\end{equation}%
where $\epsilon = 0$ corresponds to null and timelike geodesics, while $%
\epsilon = 1$ pertains specifically to timelike geodesics. The effective
potential $V_{\mathrm{eff}}\left( r\right) $, can then be defined as%
\begin{equation}
V_{\mathrm{eff}}\left( r\right) =\left( \epsilon +\frac{L^{2}}{r^{2}}\right) %
\left[ -m_{0}-\frac{R_{0}r^{2}}{6}-\frac{\eta q}{2^{1/4}\left(
1+f_{R_{0}}\right) r}\right] ,  \label{veff}
\end{equation}%
so that the radial equation takes the compact form%
\begin{equation}
\dot{r}^{2}=E^{2}-V_{\mathrm{eff}}\left( r\right) .
\end{equation}

The shape of a particle's or photon's orbit is determined by its energy $E$
and the angular momentum $L$. Since the radial coordinate $r$ must be real
and positive, the physically allowed regions correspond to values of $r$
satisfying $E^{2}\geq V_{\mathrm{eff}}\left( r\right) $, with the turning
points of an orbit occurring at $E^{2}=V_{\mathrm{eff}}\left( r\right) $.

\subsection{Timelike Geodesic Structure}

\bigskip For timelike geodesics ($\epsilon =1$), Eqs. (\ref{rad}) and (\ref%
{veff}) become%
\begin{eqnarray}
\dot{r}^{2} &=&E^{2}-\left( 1+\frac{L^{2}}{r^{2}}\right) \left[ -m_{0}-\frac{%
R_{0}r^{2}}{6}-\frac{\eta q}{2^{1/4}\left( 1+f_{R_{0}}\right) r}\right] ,
\label{Emotion} \\
&&  \notag \\
V_{\mathrm{eff}}\left( r\right) &=&\left( 1+\frac{L^{2}}{r^{2}}\right)
\left( -m_{0}-\frac{R_{0}r^{2}}{6}-\frac{\eta q}{2^{1/4}\left(
1+f_{R_{0}}\right) r}\right) .
\end{eqnarray}

\begin{figure}[]
\centering
\includegraphics[width=0.45\linewidth]{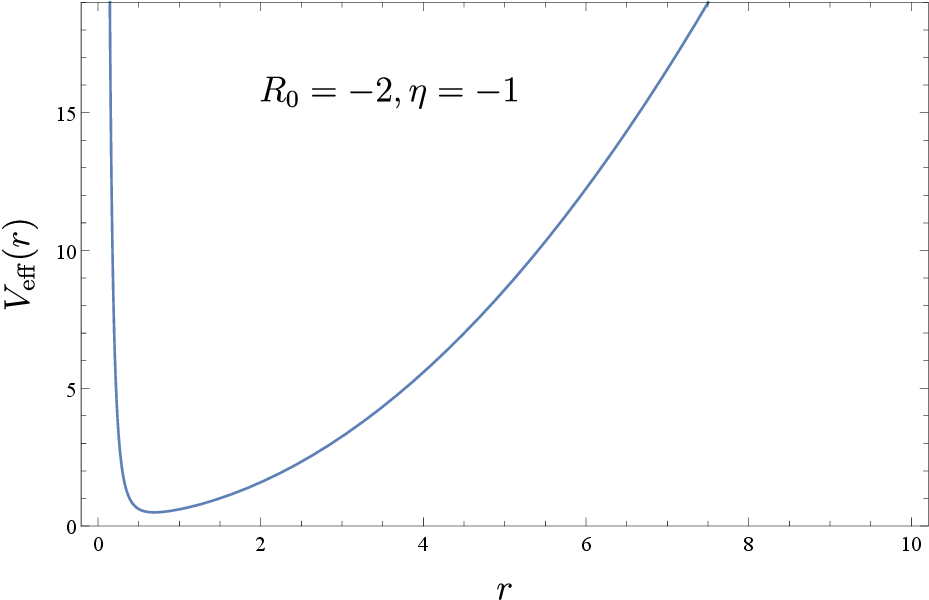} \includegraphics[width=0.45%
\linewidth]{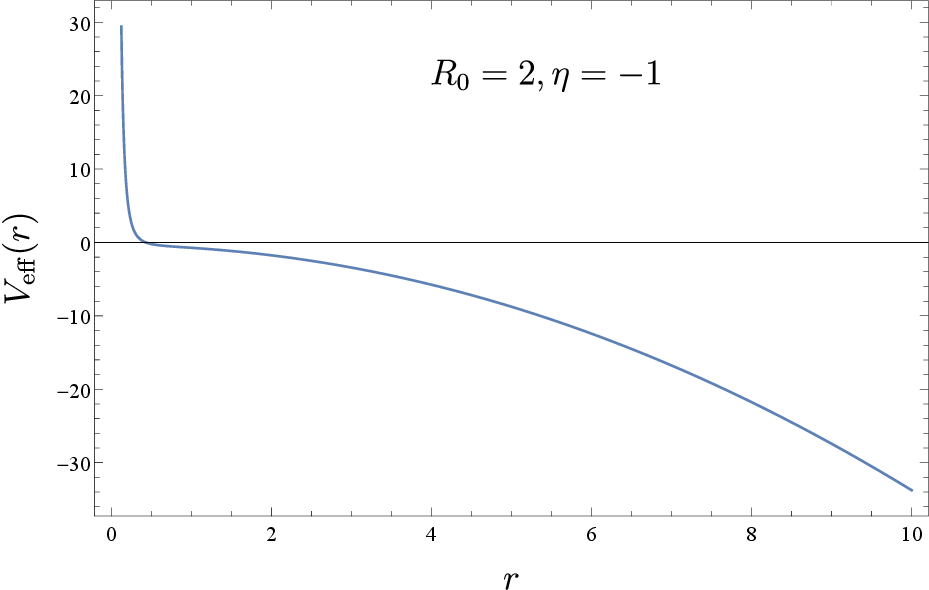} \newline
\includegraphics[width=0.45\linewidth]{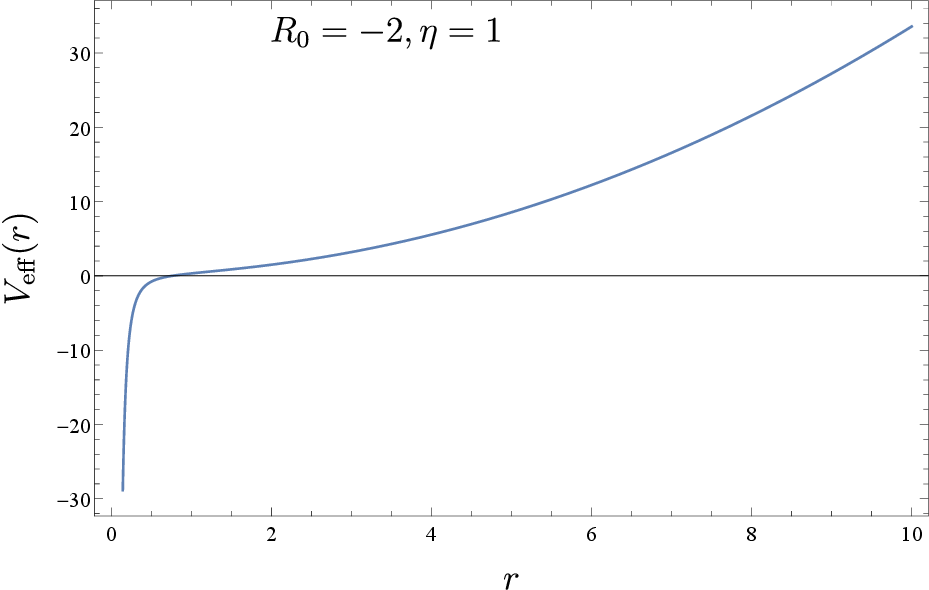} \includegraphics[width=0.45%
\linewidth]{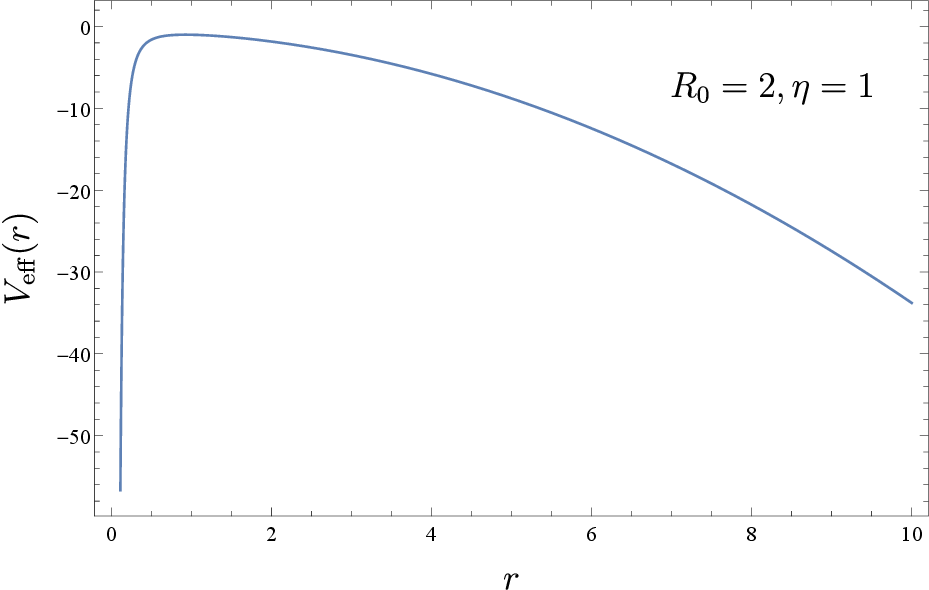} \newline
\caption{Effective potential $V_{\mathrm{eff}}\left( r\right) $ curve of
timelike geodesic}
\label{FigE4}
\end{figure}

The motion of a test particle can thus be interpreted as that of a classical
particle of energy $E^{2}$ moving in the one-dimensional potential $V_{%
\mathrm{eff}}\left( r\right) $. Figure \ref{FigE4} illustrates the effective
potential $V_{\mathrm{eff}}\left( r\right) $ for timelike geodesics in the
spacetime of the $F\left( R\right)-$BTZ black hole with parameters $%
m_{0}=0.1,$ $q=0.1,$ $L=1$ and $f_{R_{0}}=0.2$. The behavior of the
potential strongly depends on the curvature scalar $R_{0}$ and the field
parameter $\eta $. For the Maxwell configuration ($\eta =1$):

\begin{description}
\item[-] When $R_{0}<0$, the effective potential is entirely negative and
decreases monotonically with $r$, hence, no bound or stable circular orbits
are possible.

\item[-] When $R_{0}>0$, $V_{\mathrm{eff}}\left( r\right) $ starts from a
negative value near $r=0$, crosses zero, and then increases monotonically
for larger $r$, showing a repulsive behavior without any trapping region.
\end{description}

For the \textit{phantom} configuration ($\eta =-1$):

\begin{description}
\item[-] When $R_{0}<0$, the potential develops a clear minimum,
corresponding to a stable circular orbit.The surrounding potential well,
where $V_{\mathrm{eff}}\left( r\right) <E^{2}$, allows for bound,
oscillatory motion between inner and outer turning points.

\item[-] When $R_{0}>0$, the potential begins positive at small $r$, crosses
zero, and becomes negative. The initial positive barrier is repulsive, and
while the outer region is accessible, the monotonically decreasing shape
without a minimum precludes the existence of stable circular orbits.
\end{description}

Thus, Fig. \ref{FigE4} clearly shows that stable timelike circular orbits
exist only for the \textit{phantom} BTZ black hole with $R_{0}<0$, while
both the Maxwell and \textit{phantom} configurations with $R_{0}>0$ lack any
stable bound motion. This behavior signifies that the repulsive contribution
of the \textit{phantom} field can balance the gravitational attraction,
producing a trapping region where bounded motion occurs. Consequently, the
interplay between the \textit{phantom} field and the curvature correction in 
$F(R)$ gravity enhances the gravitational confinement and introduces
qualitatively new orbital structures that are absent in the standard BTZ
geometry.

The qualitative behavior of $V_{\mathrm{eff}}\left( r\right) $ determines
the possible types of orbits. The turning points of the motion correspond to
the radii where $\dot{r}=0$, that is,%
\begin{equation}
E^{2}=\left( 1+\frac{L^{2}}{r^{2}}\right) \left( -m_{0}-\frac{R_{0}r^{2}}{6}-%
\frac{\eta q}{2^{1/4}\left( 1+f_{R_{0}}\right) r}\right) .
\end{equation}

Circular orbits occur when the effective potential is at an extremum, i.e.%
\begin{equation}
\frac{dV_{\mathrm{eff}}\left( r\right) }{dr}=0.
\end{equation}

Applying this condition yields the specific angular momentum $L$\ and energy 
$E$ for circular orbits:%
\begin{eqnarray}
L^{2} &=&\frac{\left( \frac{R_{0}r^{2}}{3}-\frac{\eta q}{2^{1/4}\left(
1+f_{R_{0}}\right) r}\right) r^{2}}{\left( 2m_{0}+\frac{3\eta q}{%
2^{1/4}\left( 1+f_{R_{0}}\right) r}\right) }, \\
&&  \notag \\
E^{2} &=&\frac{-2\left( m_{0}+\frac{R_{0}r^{2}}{6}+\frac{\eta q}{%
2^{1/4}\left( 1+f_{R_{0}}\right) r}\right) ^{2}}{\left( 2m_{0}+\frac{3\eta q%
}{2^{1/4}\left( 1+f_{R_{0}}\right) r}\right) }.
\end{eqnarray}

The existence of physical timelike circular orbits requires both $L^{2}>0$
and \ \ $E^{2}>0$. These conditions lead to the following pair of
inequalities: 
\begin{equation}
2m_{0}+\frac{3\eta q}{2^{1/4}\left( 1+f_{R_{0}}\right) r}<0,~~~\&\text{ ~~~ }%
\frac{R_{0}r^{2}}{3}-\frac{\eta q}{2^{1/4}\left( 1+f_{R_{0}}\right) r}<0.
\label{58}
\end{equation}

\begin{itemize}
\item[\textbf{-}] \textbf{Maxwell field} $\eta =1:$
\end{itemize}

The first inequality $2m_{0}+\frac{3\eta q}{2^{1/4}\left( 1+f_{R_{0}}\right)
r}<0$, cannot be satisfied. Hence, no physical timelike circular orbits
exist in this case.

\begin{itemize}
\item[\textbf{-}] \textbf{Phantom} field $\eta =-1:$
\end{itemize}

In the \textit{phantom} regime, the term $\frac{3\eta q}{2^{1/4}\left(
1+f_{R_{0}}\right) r}<0$, The second inequality in Eq. (\ref{58}) then
requires $R_{0}<0$, i.e. a negative curvature background. For $R_{0}<0$, the
two inequalities combine to give the allowed range of radii for circular
orbits 
\begin{equation}
\left( \frac{3\eta q}{2^{1/4}\left( 1+f_{R_{0}}\right) R_{0}}\right) ^{\frac{%
1}{3}}<r<\frac{-3\eta q}{2^{5/4}\left( 1+f_{R_{0}}\right) m_{0}}.
\end{equation}

Therefore, timelike circular orbits exist only in the phantom regime ($\eta
=-1$) for negative curvature ($R_{0}<0$), and only if the interval in Eq. (%
\ref{58}) is non-empty 
\begin{equation}
\left( \frac{3\eta q}{2^{1/4}\left( 1+f_{R_{0}}\right) R_{0}}\right) ^{\frac{%
1}{3}}<\frac{-3\eta q}{2^{5/4}\left( 1+f_{R_{0}}\right) m_{0}}.
\end{equation}

Furthermore, the physically relevant orbits are restricted to lie outside
the event horizon, $r>r_{+}$. These conditions uniquely determine the region
in which stable circular timelike geodesics can exist around the \textit{%
phantom} BTZ black hole in $F\left( R\right) $ gravity.

The stability of circular orbits is contingent upon the sign of the second
derivative of the effective potential 
\begin{eqnarray}
\frac{d^{2}V_{\mathrm{eff}}\left( r\right) }{dr^{2}} &<&0\Longrightarrow 
\text{ Stable.} \\
&&  \notag \\
\frac{d^{2}V_{\mathrm{eff}}\left( r\right) }{dr^{2}} &>&0\Longrightarrow 
\text{ Unstable.}
\end{eqnarray}%
Thus, the minima of $V_{\mathrm{eff}}(r)$ indicate stable circular orbits,
whereas the maxima indicate unstable ones.

Table. \ref{tab:results} show some numerical values of stable circular
orbits around the \textit{phantom} BTZ black hole in $F(R)$ gravity. We
observe that the orbital radius becomes smaller as the scalar curvature $%
R_{0}$ becomes more negative. Similarly, for a fixed $R_{0}$, a larger $%
f_{R_{0}}$ also leads to a smaller orbit. This quantifies how the combined
effects of the \textit{phantom} field and the $F(R)$ gravity correction
create stable orbits that are closer to the black hole.

\begin{table}[]
\caption{ Some numerical values of stable circular orbit for different $R_0$
and $f_{R_0}$ parameters and fixed $L=1$, $m_{0}=0.1$, $q=0.1$ and $\protect%
\eta=-1$.}
\label{tab:results}\centering
\begin{tabular}{|c|c|c|c|}
\hline\hline
$f_{R_{0}}$ & {$R_{0}=-0.5$} & {$R_{0}=-1$} & {$R_{0}=-1.5$} \\ \hline\hline
0.1 & 0.920292 & 0.811588 & 0.754163 \\ \hline
0.2 & 0.880926 & 0.780872 & 0.727401 \\ \hline
0.3 & 0.843908 & 0.752111 & 0.702401 \\ \hline
0.4 & 0.808886 & 0.724992 & 0.678877 \\ \hline
0.5 & 0.775611 & 0.699281 & 0.656613 \\ \hline\hline
\end{tabular}%
\end{table}

To derive an analytic expression for the periastron advance, we must rewrite
the equation of motion (\ref{Emotion}) in terms of the angular coordinate.
For this purpose, we make use of the angular momentum relation (\ref{310})
to express the radial coordinate as a function of $\varphi $, that is $%
r=r\left( \varphi \right) $. From Eq. (\ref{Emotion}),we have%
\begin{equation}
\frac{dr}{d\tau }=\frac{dr}{d\varphi }\frac{d\varphi }{d\tau }=\frac{dr}{%
d\varphi }\frac{L}{r^{2}},
\end{equation}%
Substituting this relation into Eq. (\ref{Emotion}), we obtain%
\begin{equation}
\left( \frac{dr}{d\varphi }\right) ^{2}=\frac{R_{0}r^{6}}{6L^{2}}+\frac{%
\left( E^{2}+m_{0}+\frac{L^{2}R_{0}}{6}\right) r^{4}}{L^{2}}+\frac{\eta
qr^{3}}{2^{1/4}\left( 1+f_{R_{0}}\right) L^{2}}+m_{0}r^{2}+\frac{\eta qr}{%
2^{1/4}\left( 1+f_{R_{0}}\right) }.
\end{equation}

Then, we introduce a new variable $r=\frac{1}{u}$, and after some algebra we
can get this finale equation:%
\begin{eqnarray}
\left( u\frac{du}{d\varphi }\right) ^{2} &=&\frac{\eta qu^{5}}{2^{1/4}\left(
1+f_{R_{0}}\right) }+m_{0}u^{4}+\frac{\eta qu^{3}}{2^{1/4}\left(
1+f_{R_{0}}\right) L^{2}}+\frac{\left( E^{2}+m_{0}+\frac{L^{2}R_{0}}{6}%
\right) u^{2}}{L^{2}}+\frac{R_{0}}{6L^{2}}  \notag \\
&&  \notag \\
&=&\sum_{i=0}^{5}a_{i}u^{i},\text{ }
\end{eqnarray}%
which is a polynomial of degree 5. The solution to this equation can be
expressed in terms of higher transcendental functions, specifically the
Kleinian sigma function \cite{Hackmann,Soroushfar,Enolski}%
\begin{equation}
u\left( \varphi \right) =-\frac{\sigma _{1}}{\sigma _{2}}\left( \varphi
_{\sigma }\right) ,
\end{equation}%
where, $\sigma _{j}$ denotes the $j$-th derivative of the Kleinian sigma
function $\sigma \left( x\right) $. This function itself is defined by a
Riemann theta function with characteristic $\left[ g,h\right] $%
\begin{equation}
\sigma \left( x\right) =Ce^{x^{t}kx}\theta \left[ K_{\infty }\right] \left(
\left( 2\omega \right) ^{-1}x,\tau \right) ,
\end{equation}%
and the vector of Riemann constants with base point at infinity $\left(
0,1\right) ^{t}+\left( 1,1\right) ^{t}\tau .$ Since $r=\frac{1}{u}$, the
analytical solution for the orbital motion of massive particles is
consequently given by:%
\begin{equation}
r=-\frac{\sigma _{2}}{\sigma _{1}}\left( \varphi _{\sigma }\right) +r_{0}.
\label{33}
\end{equation}

This closed-form solution, Eq. (\ref{33}), fully describes the different
types of orbits for each region of this solution are illustrated in Fig. 2

\begin{figure}[]
\centering
\includegraphics[width=0.38\linewidth]{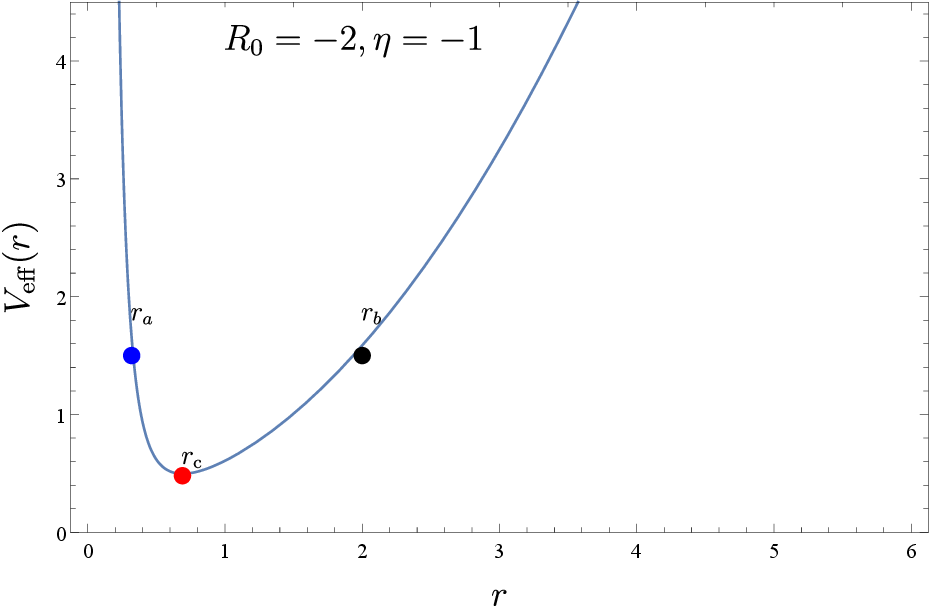} \includegraphics[width=0.26%
\linewidth]{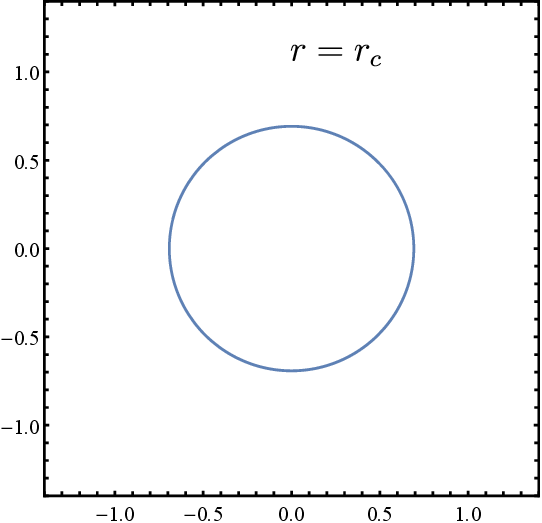} \includegraphics[width=0.254\linewidth]{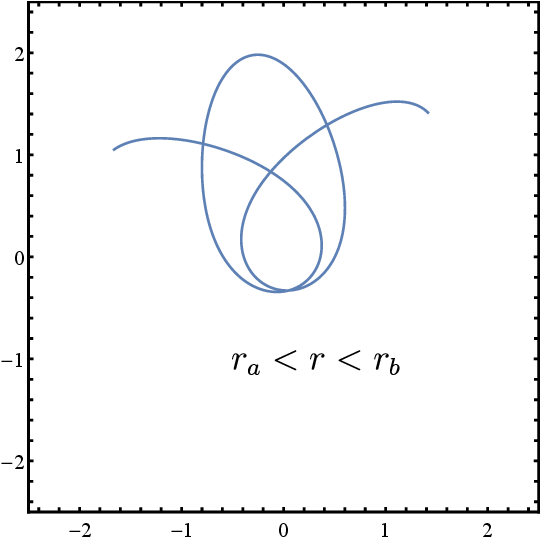} 
\newline
\caption{Effective potential curve of timelike geodesic and trajectories of
massive particle.}
\label{FigE33}
\end{figure}

\subsection{Null Geodesic Structure}

In this subsection, we analyze the motion of massless particles ($\epsilon
=0 $) in the spacetime of the \textit{phantom} BTZ black hole in $F\left(
R\right) $ gravity. For null geodesics, the effective potential is obtained
from Eq. (\ref{veff}) by setting $\epsilon =0,$%
\begin{equation}
V_{\mathrm{eff}}\left( r\right) =\frac{L^{2}}{r^{2}}\left[ -m_{0}-\frac{%
R_{0}r^{2}}{6}-\frac{\eta q}{2^{1/4}\left( 1+f_{R_{0}}\right) r}\right] .
\end{equation}

Circular photon orbits correspond to the extrema of the effective potential.
These orbits are unstable and exist at the maximum of $V_{\mathrm{eff}%
}\left( r\right) $, determined by the following conditions:%
\begin{equation}
\dot{r}=0,~~~\&\text{ ~}\frac{dV_{\mathrm{eff}}\left( r\right) }{dr}=0,\text{
\ ~}\&\text{~~~}\frac{d^{2}V_{\mathrm{eff}}\left( r\right) }{dr^{2}}<0.
\end{equation}

From the above potential, the critical impact parameter $b_{c}=\frac{L}{E}$,
is given by%
\begin{equation}
b_{c}=\frac{L}{E}=\frac{r_{\mathrm{ph}}}{\sqrt{-m_{0}-\frac{R_{0}r_{\mathrm{%
ph}}^{2}}{6}-\frac{\eta q}{2^{1/4}\left( 1+f_{R_{0}}\right) r_{\mathrm{ph}}}}%
},
\end{equation}%
where $r_{\mathrm{ph}}$ denotes the radius of the circular photon orbit.
Circular photon orbits exist only when the following reality condition is
satisfied:%
\begin{equation}
-m_{0}-\frac{R_{0}r_{\mathrm{ph}}^{2}}{6}-\frac{\eta q}{2^{1/4}\left(
1+f_{R_{0}}\right) r_{\mathrm{ph}}}>0,
\end{equation}%
This condition holds only in the \textit{phantom} regime, for suitable
choices of the parameters. It is important to note that the spacetime under
consideration is three-dimensional; therefore, no shadow area can be
defined. Instead, the critical impact parameter $b_{c}$ represents the
radius of the capture cross-section. Applying the condition $\frac{d}{dr}V_{%
\mathrm{eff}}\left( r\right) =0$ gives the radius of the circular photon
orbit:

\begin{equation}
r_{\mathrm{ph}}=-\frac{3\eta q}{2^{5/4}\left( 1+f_{R_{0}}\right) m_{0}},
\end{equation}%
Hence, the existence of a real and positive $r_{\mathrm{ph}}$ again requires
the \textit{phantom} regime. Finally, substituting $r_{\mathrm{ph}}$ into
the second derivative of $V_{\mathrm{eff}}$ yields 
\begin{equation}
\left. \frac{d^{2}V_{\mathrm{eff}}\left( r\right) }{dr^{2}}\right\vert
_{r=r_{\mathrm{ph}}}=\frac{64\left( 1+f_{R_{0}}\right) ^{4}m_{0}^{5}}{%
81q^{4}\eta ^{4}}>0.
\end{equation}

This positive value indicates that the circular photon orbit at $r_{\mathrm{%
ph}}$ is stable. Therefore, in this \textit{phantom} BTZ spacetime, circular
photon orbits exist and are stable.

Next, we consider a non-radial null geodesic, describing the motion of a
massless test particle with nonzero angular momentum. The equation of motion
for $r$ is%
\begin{equation}
\left( \frac{dr}{d\tau }\right) ^{2}=E^{2}-\frac{L^{2}}{r^{2}}\left( -m_{0}-%
\frac{R_{0}r^{2}}{6}-\frac{\eta q}{2^{1/4}\left( 1+f_{R_{0}}\right) r}%
\right) ,
\end{equation}%
or equivalently,%
\begin{equation}
\left( \frac{dr}{d\varphi }\right) ^{2}=\frac{E^{2}r^{4}}{L^{2}}-r^{2}\left(
-m_{0}-\frac{R_{0}r^{2}}{6}-\frac{\eta q}{2^{1/4}\left( 1+f_{R_{0}}\right) r}%
\right) .
\end{equation}

In the study of geodesic structure, it is convenient to introduce the
variable $u=\frac{1}{r}.$ Using this change of variable, the radial equation
becomes:%
\begin{equation}
\left( \frac{du}{d\varphi }\right) ^{2}=\left( \frac{E^{2}}{L^{2}}+\frac{%
R_{0}}{6}\right) +m_{0}u^{2}+\frac{\eta qu^{3}}{2^{1/4}\left(
1+f_{R_{0}}\right) }.  \label{cub}
\end{equation}

With the substitution 
\begin{equation}
u=\frac{\left( 4y-\frac{a_{2}}{3}\right) }{a_{3}},
\end{equation}%
the equation takes the Weierstrass form%
\begin{equation}
\left( \frac{dy}{d\varphi }\right) ^{2}=4y^{3}-g_{2}y-g_{3},  \label{cub1}
\end{equation}%
where%
\begin{eqnarray}
a_{2} &=&m_{0},~~~\&~~~a_{3}=\frac{\eta q}{2^{1/4}\left( 1+f_{R_{0}}\right) }%
, \\
g_{2} &=&\frac{m_{0}^{2}}{12},~~~\&~~~g_{3}=\frac{-\eta q^{3}\left( \frac{%
E^{2}}{L^{2}}+\frac{R_{0}}{6}\right) }{2^{19/4}\left( 1+f_{R_{0}}\right) ^{3}%
}-\frac{m_{0}^{3}}{216}.
\end{eqnarray}

Equation (\ref{cub1}) is of the elliptic type and is solved by the
Weierstrass $\wp $ function \cite{Hackmann,Soroushfar} 
\begin{equation}
y\left( \varphi \right) =\wp \left( \varphi -\varphi
_{in};g_{2};g_{3}\right) ,
\end{equation}%
with%
\begin{equation*}
\varphi _{in}=\varphi _{0}+\int_{y_{0}}^{\infty }\frac{dy}{\sqrt{%
4y^{3}-g_{2}y-g_{3}}}\text{\ },
\end{equation*}%
where \ $y_{0}=\frac{a_{3}}{4r_{0}}+\frac{a_{2}}{12}$ depends on the initial
position $r_{0}$ and $\varphi _{0}.$ Substituting back for $r,$ the solution
for the trajectory is 
\begin{equation}
r\left( \varphi \right) =\frac{\eta q}{2^{1/4}\left( 1+f_{R_{0}}\right)
\left( 4\wp \left( \varphi -\varphi _{in};g_{2};g_{3}\right) -\frac{m_{0}}{3}%
\right) }.
\end{equation}

From Eq. (\ref{cub}) the trajectory of a massless particle is fully
determined by the parameters $m_{0},q,\eta ,R_{0},f_{R_{0}}$ and the ratio $%
\frac{E}{L}$. The cubic term in $u^{3}$ arising from the \textit{phantom}
contribution $\eta q$ modifies the shape of the effective potential compared
to the standard BTZ case. Depending on the initial conditions $r_{0}$ and $%
\varphi _{0}$ the particle may spiral around the black hole in a bound-like
trajectory before escaping or falling toward the center. The Weierstrass
function solution provides a closed analytic form for these orbits,
illustrating the rich structure of geodesics in the \textit{phantom} BTZ
spacetime. This analysis highlights the significant impact of the \textit{%
phantom} parameter on the null geodesic structure.

In Fig. \ref{FigE3}, we plot the effective potential and photon trajectories
for the parameters $m_{0}=0.1$, $q=0.1$, $\eta =-1$, $f_{R_{0}}=0.2$, and $%
R_{0}=-2$, with various points marked.the effective potential develops a
pronounced minimum, indicating the possibility of stable circular orbits. In
the \textit{phantom} regime the effective potential develops a pronounced
minimum, indicating the possibility of stable circular orbits. In contrast,
for the Maxwell case, the potential is monotonic and no bounded or stable
trajectories exist.

\begin{description}
\item[- ] When $E=E_{c}$: the photon moves in a stable circular orbit with
radius $r_{E_{c}}$.

\item[- ] When $r_{B}<r<r_{A}$: the photon moves in a bound orbit within the
radial range $r_{B}<r<r_{A}$, where $r_{B}$ is the periastron and $r_{A}$ is
the apastron. 
\begin{figure}[]
\centering
\includegraphics[width=0.38\linewidth]{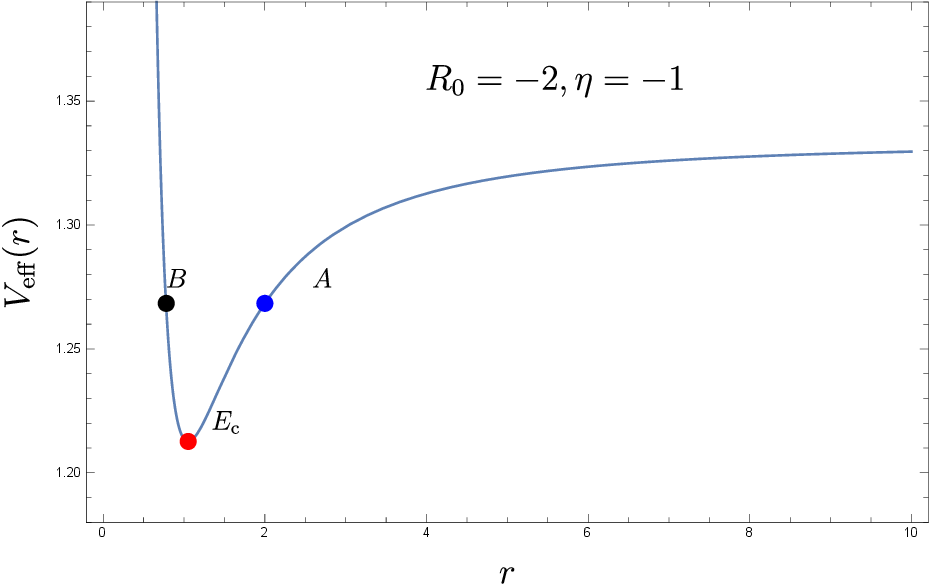} \includegraphics[width=0.26%
\linewidth]{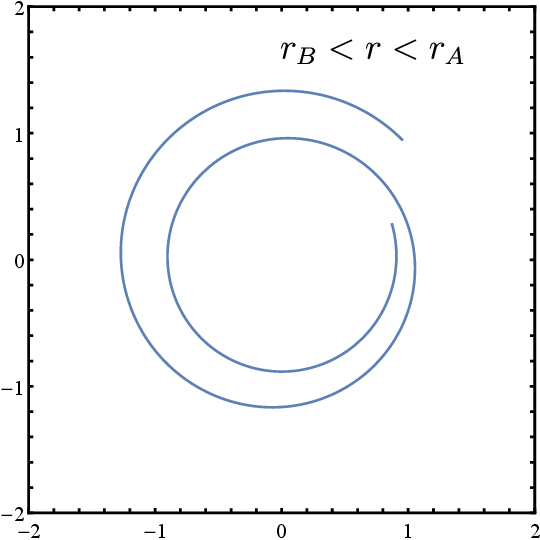} \includegraphics[width=0.27\linewidth]{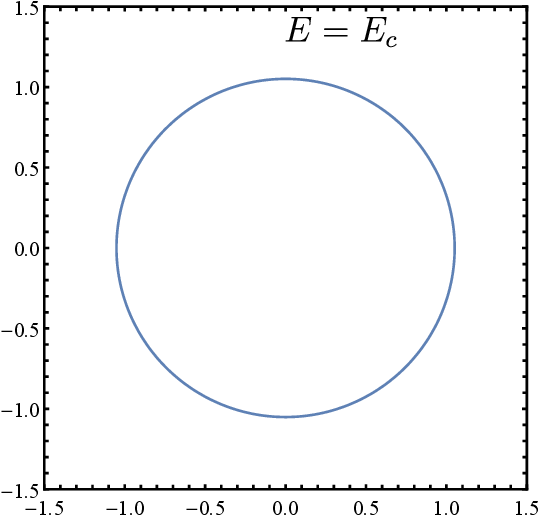} \newline
\caption{Effective potential curve of null geodesic and trajectories of
photon.}
\label{FigE3}
\end{figure}
\end{description}

\section{\textbf{Conclusions}}

We first introduced a theory of gravity that combines $F(R)$ gravity with 
\textit{power-Maxwell} theory. We derived exact solutions within the
framework of \textit{$F(R)-$conformally invariant Maxwell} theory, which
encompasses both Maxwell and \textit{phantom} fields. These solutions
imposed a constraint on the parameter $f_{R}$, indicating that $f_{R}\neq -1$%
. Next, we evaluated the Kretschmann scalar to identify the singularities of
the solutions and found a singularity at $r=0$. We analyzed the solutions to
determine the event horizon, as shown in Fig. \ref{Fig1}. Our analysis
revealed that the solutions presented in Eq. (\ref{g(r)F(R)}) could exhibit
an event horizon when $R_{0}$ is negative (i.e., $R_{0}<0$). Furthermore, we
discovered that Maxwell BTZ black holes in $F(R)$ gravity possessed only one
real root, corresponding to the event horizon. In contrast, \textit{phantom}
BTZ black holes had two real positive roots. Additionally, for the same
parameter values, Maxwell BTZ black holes were larger than \textit{phantom}
BTZ black holes.

We computed the conserved and thermodynamic quantities for the Maxwell and 
\textit{phantom} BTZ black hole solutions within the framework of $F(R)$
gravity to validate the first law of thermodynamics. Notably, our analysis
revealed that the total mass of \textit{phantom} BTZ black holes in $F(R)$
gravity was always positive. In contrast, the total mass of large Maxwell
BTZ black holes was positive, indicating that these large Maxwell BTZ black
holes are physical objects.

We analyzed black holes as thermodynamic systems to examine their local and
global stabilities through heat capacity and Gibbs potential. Our
investigation focused on the impact of the constant scalar curvature ($R_{0}$%
) and the parameter ($\eta$) on the local and global stabilities of Maxwell
and \textit{phantom} BTZ black holes within $F(R)$ gravity. The findings
regarding heat capacity, as illustrated in Fig. \ref{Fig3}, are as follows:

1- For $R_{0}>0$: Neither Maxwell ($\eta =+1$) nor \textit{phantom} ($\eta
=-1$) BTZ black holes met the criteria for physical stability when $R_{0}>0$%
. Specifically, both temperature and heat capacity were not positive at the
same time, indicating that these black holes could not exist.

2- For $R_{0}<0$: Maxwell ($\eta =+1$) and \textit{phantom} ($\eta =-1$) BTZ
black holes with large entropy (or large radius) proved to be physical and
stable. Notably, we observed two distinct behaviors between the two types of
black holes: i) A phase transition point was identified between small
unstable and large stable Maxwell BTZ black holes, whereas a physical
limitation point was found between the small unstable and large stable 
\textit{phantom} BTZ black holes. ii) The region of physical stability for 
\textit{phantom} BTZ black holes was larger than that for Maxwell BTZ black
holes.

We evaluated the global stability of Maxwell and \textit{phantom} BTZ black
holes within the $F(R)$ framework by applying the Gibbs potential, which we
plotted in Fig. \ref{Fig4}. Our analysis revealed that larger Maxwell BTZ
black holes are globally stable objects. In contrast, \textit{phantom} BTZ
black holes consistently meet the global stability condition. Therefore, 
\textit{phantom} BTZ black holes in $F(R)$ theory of gravity are globally
stable for any radius.

We investigated the volume expansion coefficient ($\alpha$), the isothermal
compressibility coefficient ($\kappa_{T}$), and the specific heat at
constant pressure ($C_{P}$). Our findings revealed that all three quantities
shared a common factor in their denominators and exhibited divergence at the
critical point. We conducted an analytical verification of the Ehrenfest
equations, confirming that both relations were satisfied. Furthermore, we
calculated the Prigogine-Defay ratio, which was determined to be exactly
equal to one. Consequently, we characterized the phase transition at the
critical point as a second-order transition.

Furthermore, we investigated the geodesic structure of the obtained
spacetime to understand the influence of the \textit{phantom} field and the $%
F(R)$ correction on the motion of test particles. Our analysis revealed that
stable timelike circular orbits exist only in the \textit{phantom} regime ($%
\eta=-1$) for negative curvature backgrounds, whereas no bound or stable
motion is possible in the Maxwell case. For null geodesics, we found that
the \textit{phantom} BTZ configuration admits stable circular photon orbits
characterized by a real and positive photon radius, while the corresponding
Maxwell case exhibits no such behavior. The analytical treatment of both
timelike and null trajectories demonstrated that the presence of the phantom
parameter significantly modifies the effective potential and enriches the
orbital dynamics of the system. These findings confirm that the inclusion of
the \textit{phantom} field in $F(R)$ gravity leads to qualitatively new and
physically interesting geodesic structures compared with the standard BTZ
solution.

\begin{acknowledgements}
The authors express gratitude to the esteemed referee for the valuable comments. This research was financed by a research grant from the the University of Mazandaran. 
\end{acknowledgements}

\end{document}